\documentclass[12pt,preprint]{aastex}

\newcommand{\feh}{\mbox{${\rm[Fe/H]}$}}
\newcommand{\afe}{\mbox{${\rm[\alpha/Fe]}$}}
\newcommand{\mfe}{\mbox{${\rm[m_i/Fe]}$}}
\newcommand{\ofe}{\mbox{${\rm[O/Fe]}$}}
\newcommand\msol{{\cal M_{\odot}}}

\newcommand\teff{{T_{\rm eff}}}
\newcommand\amlt{{\alpha_{\rm MLT}}}

\newcommand\lta{\mathrel{\hbox{\raise 0.6 ex \hbox{$<$}\kern
                   -1.8 ex\lower .5 ex\hbox{$\sim$}}}}
\newcommand\gta{\mathrel{\hbox{\raise 0.6 ex \hbox{$>$}\kern
                   -1.7 ex\lower .5 ex\hbox{$\sim$}}}}
\newsavebox\uscorebox
 
\shortauthors{VandenBerg et al.}
\shorttitle{Victoria-Regina Isochrones}
 
\begin{document}
 
\title{ISOCHRONES FOR OLD ($> 5$ GYR) STARS AND STELLAR 
POPULATIONS.~I.~MODELS FOR $-2.4 \le$ [Fe/H] $\le +0.6$, $0.25 \le Y \le 0.33$,
AND $-0.4 \le$ [$\alpha$/Fe] $\le +0.4$}

\author{Don A.~VandenBerg}
\affil{Department of Physics \& Astronomy, University of Victoria,
       P.O.~Box 1700 STN CSC, Victoria, B.C., V8W~2Y2, Canada}
\email{vandenbe@uvic.ca}

\author{Peter A.~Bergbusch}
\affil{Department of Physics, University of Regina, Regina, Saskatchewan,
       S4S~0A2, Canada}
\email{pbergbusch@accesscomm.ca}

\author{Jason W.~Ferguson}
\affil{Department of Physics, Wichita State University, Wichita KS 67260-0032,
       U.S.A.}
\email{Jason.Ferguson@wichita.edu}

\author{Bengt Edvardsson}
\affil{Uppsala Astronomical Observatory, Department of Physics \& Astronomy,
       Uppsala University, Box 516, SE-751~20 Uppsala, Sweden}
\email{Bengt.Edvardsson@physics.uu.se}

\begin{abstract}
Canonical grids of stellar evolutionary sequences have been computed for the
helium mass-fraction abundances $Y = 0.25$, 0.29, and 0.33, and for iron
abundances that vary from $-2.4$ to $+0.4$ (in 0.2 dex increments) when
[$\alpha$/Fe] $= +0.4$, or for the ranges $-2.0 \le$ [Fe/H] $\le +0.6$, $-1.8
\le$ [Fe/H] $\le +0.6$ when [$\alpha$/Fe] $= 0.0$ and $-0.4$, respectively. The
grids, which consist of tracks for masses from $0.12 \msol$ to 1.1--$1.5
\msol$ (depending on the metallicity) are based on up-to-date physics,
including the gravitational settling of helium (but not metals diffusion).
Interpolation software is provided to generate isochrones for arbitrary ages
between $\approx 5$\ and 15 Gyr and any values of $Y$, [$\alpha$/Fe], and
[Fe/H] within the aformentioned ranges.  Comparisons of isochrones with
published color-magnitude diagrams (CMDs) for the open clusters M$\,$67 ([Fe/H]
$\approx 0.0$) and NGC$\,$6791 ([Fe/H] $\approx 0.3$) and for four of the
metal-poor globular clusters (47 Tuc, M$\,$3, M$\,$5, and M$\,$92) indicate
that the models for the observed metallicities do a reasonably good job of
reproducing the locations and slopes of the cluster main sequences and giant
branches.  The same conclusion is reached from a consideration of plots of
nearby subdwarfs that have accurate {\it Hipparcos} parallaxes and metallicities
in the range $-2.0 \lta$ [Fe/H] $\lta -1.0$ on various CMDs and on the
$(\log\teff,\,M_V)$-diagram.  A relatively hot temperature scale similar to
that derived in recent calibrations of the infrared flux method is favored by
both the isochrones and the adopted color transformations, which are based on
the latest MARCS model atmospheres.
\end{abstract}
 

\section{Introduction}
\label{sec:intro}

One of the defining properties of globular clusters (GCs) is that they show
star-to-star differences in the abundances of the light elements, as manifested
by C--N, O--Na, and (in some cases) Mg--Al anticorrelations (\citealt{cbg10};
the recent review by \citealt{gcb12}).  With relatively few exceptions (notably
$\omega$~Cen, e.g., \citealt{jp10}; and M$\,$22, \citealt{mmp09}), the [Fe/H]
values of member stars do not vary by more than a few hundredths of a dex, if
that (though some variations with evolutionary state, primarily in the vicinity
of the turnoff (TO), are expected to be the consequence of atomic diffusion ---
see, e.g., \citealt{gkr13}).  However, in at least a few systems, there are
strong indications that the helium and/or the total C$+$N$+$O abundances vary
significantly; see the studies of NGC$\,$2808 by \citet{pba07} and of
NGC$\,$1851 by \citet{mvp08}.  More commonly deduced are helium mass-fraction
abundance variations amounting to $\delta Y \lta 0.03$ (e.g., \citealt{dsf13},
\citealt{ngp13}, \citealt{gls13}).  In addition, it is known that clusters of
the same [Fe/H] can have quite different abundances of the so-called
``$\alpha$-elements" (O, Ne, Mg, Si, S, Ar, Ca, and Ti).  Whereas most of the
Milky Way GCs with [Fe/H] $\lta -0.8$ appear to have [$\alpha$/Fe] $\approx
0.35$ (\citealt{cbg09b}), a value near 0.0 has been derived for a few of them,
including, e.g., Palomar 12 (\citealt{coh04}), which seems to be connected to
the Sagittarius dwarf galaxy (\citealt{dmg00}).

To interpret photometric data for the ancient stellar populations found in GCs
--- or those residing in, e.g., nearby dwarf galaxies or the Galactic Bulge,
which are characterized by different relations between [$\alpha$/Fe]
and [Fe/H], as well as other chemical peculiarities (see, e.g., \citealt{vis04},
\citealt{lhz07}, \citealt{rge10}) --- it is obviously important to use stellar
models for the {\it observed} abundances.  It has long been known, for instance,
that the locations on the H-R diagram of the main-sequence (MS) and red-giant
branch (RGB) segments of isochrones for a fixed age ($\gta 8$ Gyr) and low
metallicities depend on $Y$, but not their subgiant branches (SGBs), which are
predicted to be nearly coincident except for a small change in slope
(\citealt{car81}).  On the other hand, the CNO elements mainly affect the
luminosity of the TO and the SGB, with little or no impact on the lower MS
or the RGB (\citealt{bv92}).  Furthermore, as reported by \citet[hereafter
V12]{vbd12}, the temperatures of red giants appear to be controlled chiefly by
Mg, Si, and Fe, which are the most abundant metals that are also important
electron donors.  Among the ground-breaking papers that have examined the
effects on stellar models of varying the $\alpha$-element abundances are those
by \citet{scs93}, \citet{vsr00}, and \citet{pcs06}, whereas the implications for
observed color-magnitude diagrams (CMDs) of light-element anticorrelations have
been studied by \citet{swf06} and \citet[\citealt{cmp13}]{csp08}.

The main purpose of this investigation is to present updated Victoria-Regina
evolutionary tracks and isochrones that allow for variations in [$\alpha$/Fe],
[Fe/H], and $Y$.  The next paper in this series will provide isochrones in
which [O/Fe] is also treated as a free parameter, whereas subsequent studies
will assess the implications of $\pm 0.2$ dex variations in [Mg/Fe] and
[Si/Fe] over a wide range in [Fe/H].  Once these projects have been
completed, it will be possible, using the interpolation codes provided in
this study (see the Appendix), to generate evolutionary tracks and isochrones
for metals mixtures that have arbitrary amounts (within reason) of O, Ne, Mg,
and Si at [Fe/H] values of interest, assuming (at least initially)
[$m$/Fe] $= 0.0$ and $+0.4$ for the other $\alpha$-elements.  

There are two main differences between our computations and the model grids
reported by other groups (e.g., \citealt{pcs06}, \citealt{dcj07},
\citealt{bmg12}, \citealt{dvd12}).  First, as described in \S~\ref{sec:abund}
(following a brief overview of the Victoria stellar structure code in
\S~\ref{sec:input}), our tracks have been calculated for specific values of
[Fe/H], instead of $Z$ (the total mass-fraction abundance of the metals), in
order to facilitate direct comparisons between theory and observations.
(At constant $Z$, the [$m$/H] value of {\it each} metal changes
when the $\log N$ abundance of any one of the heavy elements is modified in the
assumed metals mixture.  For this reason, the sensitivity of stellar models to
the abundances of individual metals, or groups of metals, should be inferred
from computations at constant [Fe/H], as in the study by V12, rather than those
generated for fixed values of $Z$; see, e.g., \citealt{dcf07}.)

Second, the accompanying software, which is briefly described in the Appendix,
interpolates simultaneously in all three of the chemical abundance parameters
so that isochrones may be generated for arbitrary values of [$\alpha$/Fe],
[Fe/H], and $Y$ (within the ranges encompassed by the models), as well as age.
As a result of this flexibility, the consequences of differences in the assumed
helium and/or $\alpha$-element abundances for observed CMDs, the derived
mass-radius diagrams of binary stars, etc., may be readily evaluated.  Indeed,
one may also examine the implications of different relations between $Y$ and/or
[$\alpha$/Fe] with [Fe/H].  In \S~\ref{sec:vrmod}, plots of selected isochrones
on the H-R diagram are presented and discussed, while a few examples of the
application of our isochrones to observational data are provided in
\S~\ref{sec:obs}.  Brief concluding remarks are given in \S~\ref{sec:sum}.

\section{The Victoria Stellar Evolution Code}
\label{sec:input}

The evolutionary code described in considerable detail by V12 has been used
to compute all of the stellar models reported in this paper.  In all respects,
up-to-date physics and, in particular, a careful treatment of the gravitational
settling of helium has been incorporated into it, together with sufficient
extra mixing below envelope convection zones (when they are present) to satisfy,
in particular, the solar Li abundance constraint.  The settling of the metals
has not been considered, but the neglect of this physics is largely
inconsequential, since a change in the central CNO abundances at the several
percent level, arising solely from diffusive processes, will have little effect
on TO luminosity versus age relations.  (Because the effective temperature
of the turnoff will be altered to some extent by the diffusion of the metals,
this physics will affect the {\it absolute magnitude} of the TO by a small
amount through the bolometric corrections.)  This assertion is supported by the
fact that the abundances of the CNO elements must be increased by about a factor
of two (i.e., $\sim 15$--20 times the enhancement caused by settling) in the
nuclear-burning region of a star in order to reduce the predicted age at a given
turnoff luminosity by 1 Gyr (see V12).  Most of the $\sim 10$\% reduction in
age that is generally attributed to the inclusion of atomic diffusion
is therefore due to the settling of helium over the star's core H-burning
lifetime (\citealt{pv91}; \citealt{ccd97}).
 
To be sure, models that take the diffusion of the metals into account predict
that the transition from lower-mass stars that possess radiative cores at
central H exhaustion to those of higher mass that have convective cores at the
end of the MS phase occurs at a lower mass and luminosity than when
this physics is not treated (\citealt{mrr04}).  However, the relevant masses
($\gta 1.1 \msol$, depending on the metallicity) evolve to the RGB tip in less
than $\sim 5$ Gyr, whereas the focus of the present study is on stars with
longer lifetimes.  Moreover, both settling and radiative accelerations should
be treated in order to provide the best possible predictions of the chemical
abundances at the surfaces of stars as a function of their evolutionary state
(see \citealt[and references therein]{mrr10}).  On the other hand, most
spectroscopic studies have failed to detect any differences in the surface
metallicities of GC stars between the TO and the lower RGB (e.g.,
\citealt{gbb01}, \citealt{rc02}) --- unless a hot effective temperature
($\teff$) scale is assumed (\citealt{kgr07}, \citealt{gkr13}); but even then,
the observed [$m$/H] variations ($\lta 0.15$ dex) are considerably less than
expected.  Improved consistency with the model predictions can be obtained if
an {\it ad hoc} additional mixing process, perhaps due to turbulence, is assumed
to occur at the bottom of surface convection zones when they are present
(\citealt{rmt00}; \citealt{rmr01}), but current formulations of this extra
mixing (also see V12) involve free parameters that must be calibrated using
observations (e.g., Li abundance data). 

Insofar as the calculation of isochrones is concerned, the
neglect of metals diffusion is not a serious omission because this physics
mainly affects the predicted temperatures of stars (in a relatively minor way;
see V12, their Fig.~1), which are subject to many other uncertainties (see
\S~\ref{sec:obs}).  In particular, the surface boundary conditions play a major
role in determining the model $\teff$\ scale. 

\subsection{The Atmospheric Boundary Conditions}
\label{subsec:atmbc}

To model the lowest masses (each grid has a minimum mass of $0.12 \msol$),
MARCS model atmospheres (\citealt{gee08}) at an optical depth $\tau = 100$ were
attached to the interior structures following
the procedures described by \citet{vee08}.  In the case of 0.3--$0.4 \msol$
models (or somewhat higher masses in the metal-rich or super-metal-rich
regimes), the stellar photosphere was taken to be the outer boundary and the
pressure at $T = \teff$ was determined by integrating the hydrostatic equation
from very small optical depths to the photospheric value, assuming the
semi-empirical \citet[hereafter HM74]{hm74} $T$--$\tau$ structure (specifically,
the fit to the latter given by \citealt{vp89}).  [When using the Sun to
calibrate such quantities as the convective mixing-length parameter, it is
obviously important to assume the {\it solar} atmospheric structure instead of,
say, a grey atmosphere (see, e.g., \citealt{mvp94}).  Encouragingly, the
temperature stratification predicted by recent 3D model atmospheres for solar
parameters appears to satisfy observational constraints even better than the
HM74 model, which is preferable to current 1D model atmospheres in representing
the surface layers of the Sun; see the discussion by \citealt{pac13}.]

\begin{figure}
\plotone{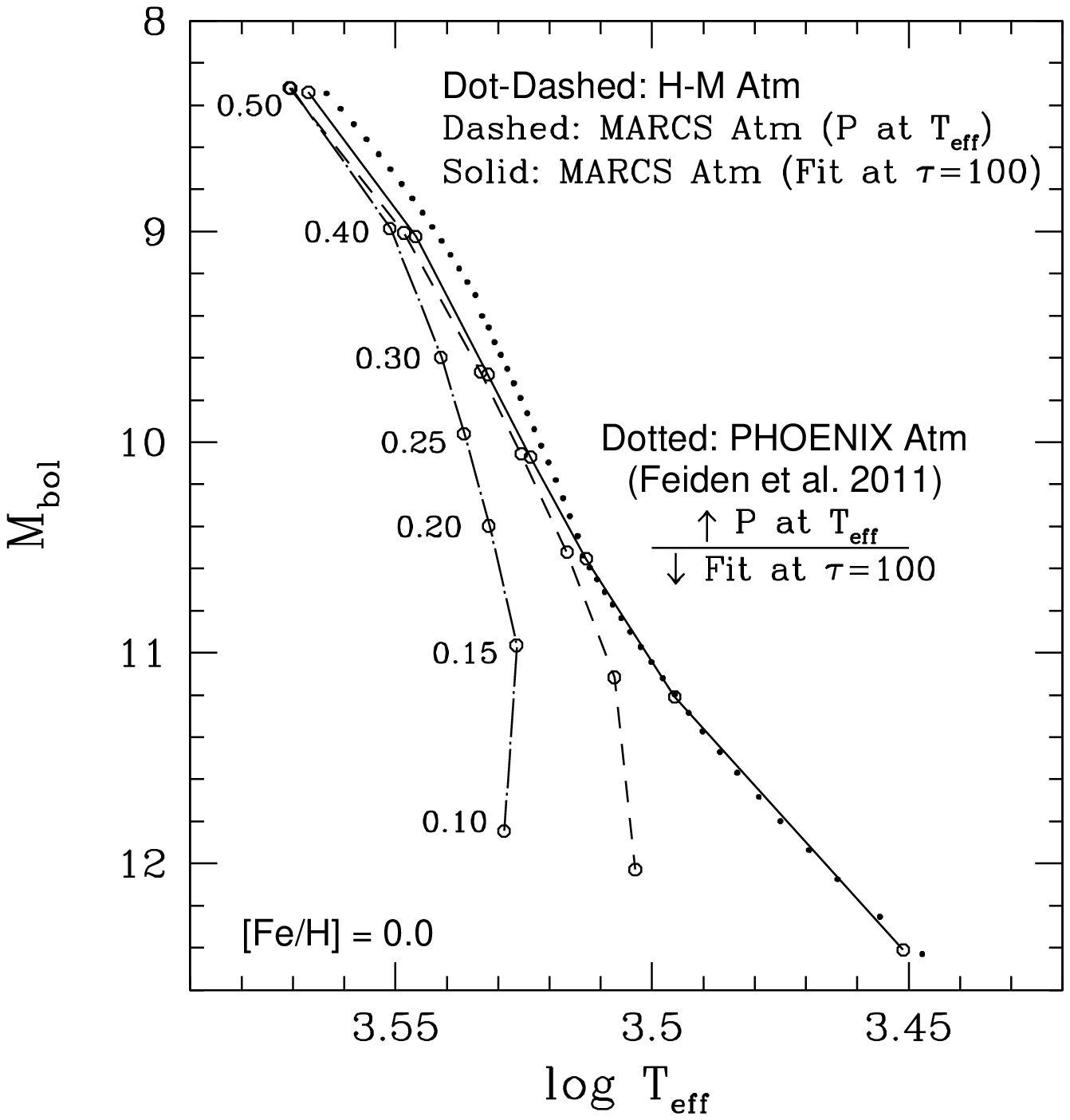}
\caption{Plot on an H-R diagram of 0.10 to $0.50 \msol$ models (open circles)
for [Fe/H] $= 0.0$ and an age of 5 Gyr, on the assumption of three different
treatments of the atmospheric layers.  The dashed and solid curves connect the
models that were obtained by fitting MARCS model atmospheres (\citealt{gee08})
to the interior structures at the photosphere (the layer where $T = \teff$) or
at $\tau = 100$, respectively.  The scaled-solar, \citet{hm74} $T$--$\tau$
relation was used to derive the boundary pressure (see the text) in the case of
the models that are connected by the dot-dashed curve.  The small filled circles
represent the lower-MS segment of a 4 Gyr, solar-metallicity isochrone provided
by G.~Feiden (2011, priv.~comm.).  PHOENIX model atmospheres (\citealt{haf99})
were used to provide the boundary conditions for these models, with the fitting
point chosen to be photosphere or $\tau = 100$ for  masses $> 0.2 \msol$ or
$\le 0.2 \msol$, respectively.}
\label{fig:fig1}
\end{figure}

The ramifications of different treatments of the atmospheric layers for
low-mass, [Fe/H] $= 0.0$ stellar models are shown in Figure~\ref{fig:fig1}.
(For an instructive example of similar work carried more than 15 years ago, see
\citealt{bcc98}.)  Using the properties of MARCS model atmospheres at
$\tau = 100$ to derive the outer boundary conditions (BCs) of stellar models
clearly results in increasingly cooler temperatures and reduced luminosities
with decreasing mass than attaching the same atmospheres to the interior
structures at $T = \teff$ or deriving the boundary pressure using the scaled
HM74 $T$--$\tau$ relationship (compare the locations of the open circles along
the solid, dashed, and dot-dashed curves, respectively).  For these
computations (and, indeed, for all of the lower-MS models in which the MARCS
atmospheres were attached at depth), the advanced equation-of-state (EOS)
developed by A.~Irwin was used in its most efficient ``EOS4" mode.\footnote{See
http://freeeos.sourceforge.net}  Because Irwin's EOS, even in the EOS4 mode, is
slower by a factor of 3--4 than the EOS which is normally employed by the
Victoria code (see \citealt{vsr00}), we have opted to use the latter for higher
mass models (where the tracks are essentially independent of this choice, see
V12). 
 
Since the lowest mass models are based a different EOS and surface BCs
than those for higher masses, it is necessary to make a smooth transition
between the two regimes.  To accomplish this, evolutionary tracks for masses in
the range 0.3--$0.8 \msol$ were computed (for each combination of the chemical
abundance parameters) to just past the zero-age main-sequence (ZAMS)
location on the H-R diagram, assuming the EOS and atmospheric BCs employed in
the lower-main-sequence (LMS) grids, on the one hand, and those used in the
higher mass tracks, on the other.  The mass for which the  differences in
$\log\,L/L_\odot$ and $\log\,\teff$ of the respective ZAMS models were the
smallest was taken to be the transition mass, and the mean luminosity and
temperature differences, which were usually $< 0.003$ in both $\log\,L/L_\odot$
and $\log\,\teff$, were calculated.  These offsets were applied to a complete
track for the transition mass (assuming the physics that has been employed for
higher masses), which became the adopted track for that mass.  No adjustments
of any kind were made to the evolutionary sequences for lower or higher masses.  

\begin{figure}
\plotone{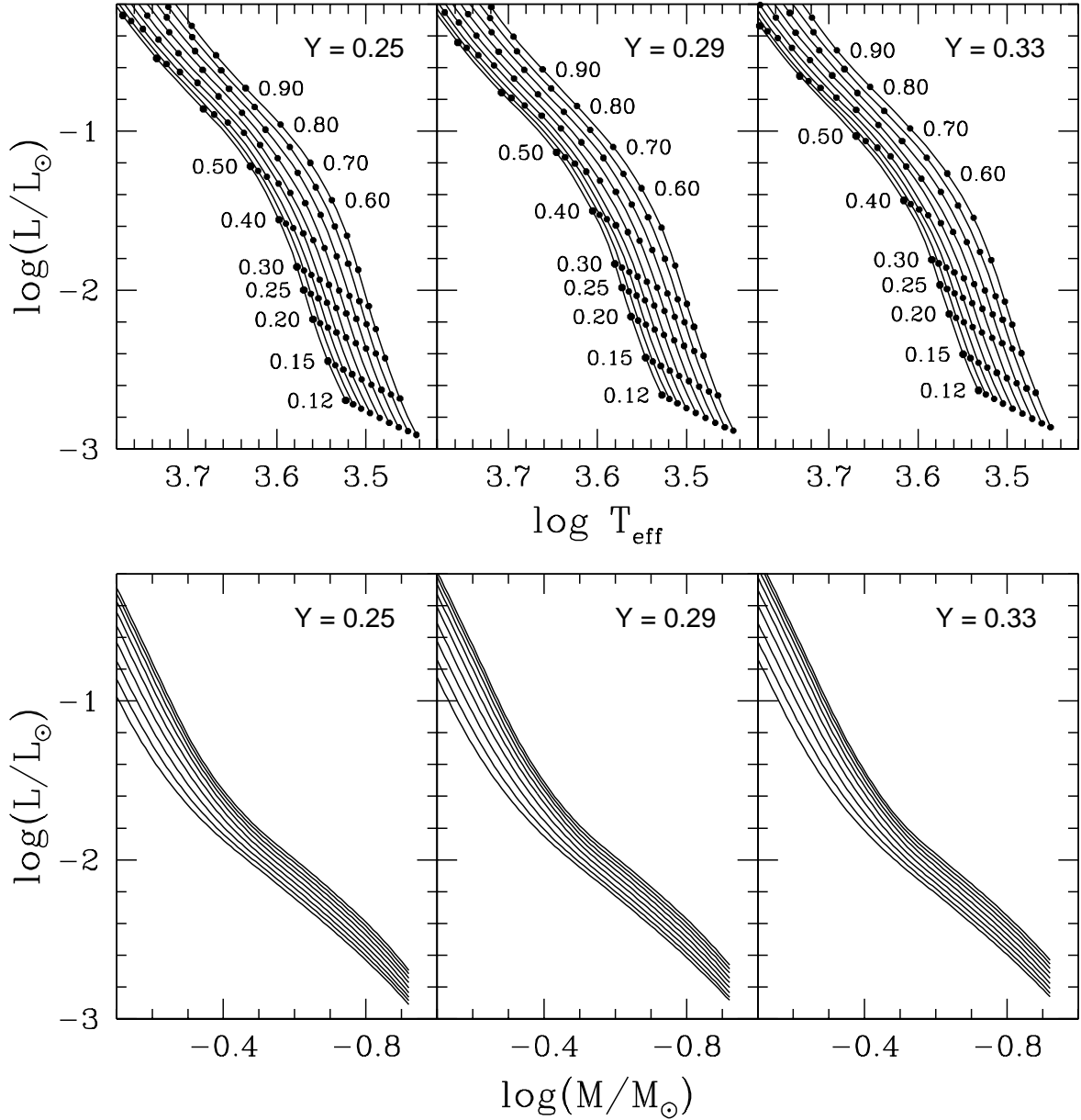}
\caption{{\it Upper panels}: Plot on the H-R diagram of cubic spline fits
(solid curves) to the ZAMS locations (filled circles) of models for the
indicated masses and helium abundances, assuming [$\alpha$/Fe] $= 0.0$ and
[Fe/H] values ranging from $-1.0$ to $+0.6$, in 0.2 dex increments (in the
direction from left to right).  At a given mass, the ZAMS models move along
diagonal lines toward lower luminosities and cooler temperatures as the
metallicity increases: note the progression of the filled curcles that
represent each mass value.  {\it Lower panels}: Plot of the same loci that
are shown in the upper panels, except on the mass-luminosity plane.}
\label{fig:fig2}
\end{figure}

As shown in Figure~\ref{fig:fig2} for a subset of the model grids, the
resultant H-R diagrams (upper panels) and mass-luminosity relations (lower
panels) are very smooth in the LMS region where this join has been made (and
elsewhere).  Only in magnified versions of this and similar plots are some
{\it very slight} irregularities evident; e.g., the spacing between the
fourth and fifth loci close to the $0.4 \msol$ ZAMS models in the upper
panel for $Y = 0.33$ is a bit larger than those between the third and fourth or
fifth and sixth loci.  However, they have no obvious impact on the predicted
mass-luminosity relations.  Note that some kind of a transition is unavoidable
because proper model atmospheres, attached at depth, must be used as boundary
conditions for very low mass models in order to obtain the most realistic
$\teff$ scale, whereas scaled HM74 atmospheric structures are the preferred
choice for the Sun and solar-type stars (as discussed below).

Fig.~\ref{fig:fig1} also plots the lower-MS portion of a 4 Gyr isochrone for
solar abundances (the dotted curve) that was kindly provided to us by G.~Feiden
(2011, private communication).  The BCs for these models (see \citealt{fcd11})
were derived from PHOENIX model atmospheres (\citealt{haf99}): the latter were
attached to the interior structures at $\tau = 100$ if ${\cal M} \le 0.2 \msol$,
or at the photosphere, in the case of higher masses.  Remarkably, the lowest
mass models overlay those represented by the solid curve nearly perfectly,
while at ${\cal M} \gta 0.2 \msol$, they are cooler than our
atmosphere-interior models by $\lta 50$\ K.  This is really quite good
agreement between completely independent predictions of the properties of very
low mass stars.  

\begin{figure}
\plotone{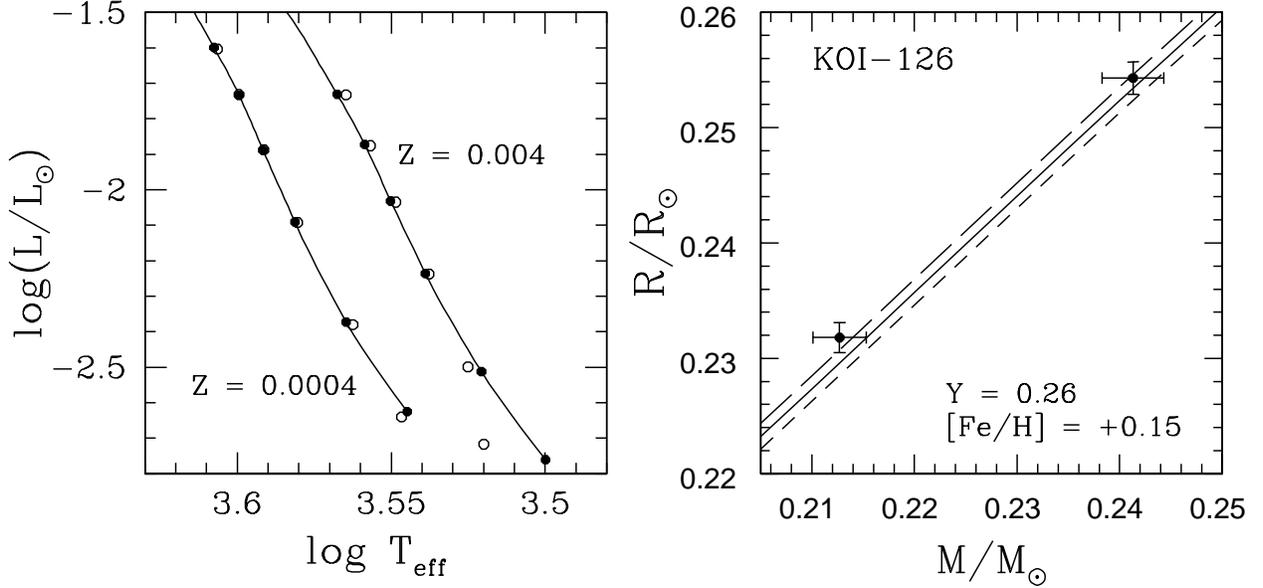}
\caption{{\it Left-hand plot}: the solid curves represent 10 Gyr isochrones
for the indicated values of $Z$ (from our grids for [$\alpha$/Fe] $= 0.4$ and
$Y = 0.25$), while the filled circles along them give the locations of 0.12,
0.15, 0.20, 0.25, 0.30, and $0.35 \msol$ models (in the direction from lower to
higher luminosities).  The open circles indicate the locations of models for 
the same $Z$, [$\alpha$/Fe], and masses (assuming $Y = 0.245 + 1.6\,Z$) along 
isochrones (not shown) computed by \citet{dcj07}.  (It is not known why the
faintest open circle for $Z=0.004$ deviates so much from our model for the same
mass and $Z$.)  {\it Right-hand plot}: Comparison on the mass-radius plane of a
5 Gyr isochrone (solid curve) for the indicated values of $Y$ and [Fe/H] with
the observed properties of the lowest mass components of the triple system
KOI-126 (Carter et al.~2011, Feiden et al.~2011).  The assumed value of $Y$ is
just slightly larger than the value required by a Standard Solar Model
($Y = 0.2553$).  The short-dashed and long-dashed curves represent isochrones
that are otherwise the same, except that they assume [Fe/H] $= +0.07$ and
$= +0.23$, respectively, to illustrate the effects on the mass-radius diagram
of varying the observed metallicity by $\pm 1\,\sigma$.}
\label{fig:fig3}
\end{figure}

At lower metallicities, the differences between the present Victoria-Regina
and published Dartmouth models (\citealt{dcj07}) are even less, as shown in
the left-hand plot in Figure~\ref{fig:fig3}.  Although they employ different
model atmospheres as boundary conditions, the latter are apparently sufficiently
similar to MARCS atmospheres that they yield nearly the same $\teff$\ scale.
Moreover, it is apparent from the near coincidence of the filled and open
circles that the predicted mass-luminosity relations (especially at masses
$> 0.15 \msol$) are in
excellent agreement as well.  However, perhaps the most compelling demonstration
of the reliability of our computations for LMS stars is provided in the
right-hand plot, which illustrates how well our models reproduce the
mass-radius relation that describes the lowest mass (0.213 and $0.241 \msol$)
components of the triple system, KOI-126 (\citealt{cfr11}).  A similar (equally
successful) comparison, but using Dartmouth models, was reported by Feiden et
al.~(2011), who note that some low-mass binaries (notably CM Draconis) continue
to be problematic, possibly due to the neglect of magnetic fields and activity
effects (also see \citealt{fc13}).  Regardless, Fig.~\ref{fig:fig3} and 
additional plots presented later in this paper (see \S~\ref{sec:obs}) provide
encouraging support for the quality of our MS and LMS models.  [Although
discrepancies between predicted and observed CMDs are generally found at the
faintest absolute magnitudes (e.g., \citealt{rdh08}, \citealt{cv14}), they
are likely due mostly to deficiencies in current color--$\teff$\ relations,
given the great difficulty of accounting for all of the sources of blanketing
in model atmospheres and synthetic spectra for cool stars.]

One question that warrants some discussion is the following: why has the scaled
solar HM74 $T$--$\tau$ relation been used to determine the boundary
pressures for the majority of our computed models instead of MARCS model
atmospheres?  As already mentioned, the former is the preferred choice for the
calculation of a Standard Solar Model, and presumably for models which are
relevant to stars with properties similar to that of the Sun.  In fact,
\citet{vee08} showed that surface BCs derived in this way agree rather well
with those obtained from scaled, differentially corrected (SDC) MARCS models
over wide ranges in $\teff$, gravity, and metallicity.  (Note that the latter
were constructed in order that the resultant solar model atmosphere reproduces
the temperature structure derived by HM74.)  While it is not necessarily the
case that the SDC atmospheres provide a better representation of those
applicable to, e.g., metal-deficient stars than standard MARCS models, the
implied $\teff$\ scale agrees quite well with that derived by \citet{crm10}
(via the infra-red flux method) for field subdwarfs that have $-2.0 \lta$ [Fe/H]
$\lta -0.6$ (see \citealt[and \S~\ref{subsec:sbd} later in this paper]{vcs10}).

As regards the \citet{vee08} examination of the use of MARCS model atmospheres
as BCs for interior structures, it is pertinent to note that an identical
treatment of convection and the same chemical abundances were assumed in the
Victoria and MARCS codes.  Indeed, plots were included in the paper by
VandenBerg et al.~to illustrate the close agreement of, among other things, the
predicted variations with depth (in the atmosphere) of the pressure, the
adiabatic temperature gradient, the opacity, and the convective flux.  It is not
possible to obtain similar consistency with the large grids of model atmospheres
published by \citet{gee08} because, for one thing, they assumed a value of 1.5
for the mixing-length parameter $\amlt$, whereas the evolutionary models
presented in this paper required $\amlt = 2.007$ to satisfy the solar
constraint.  There are also minor differences in the adopted heavy-element
abundances, which could have some effect on the opacities at low temperature.
In addition, the MARCS atmospheres were computed for a fixed helium abundance
($\log N$(He)$ = 10.93$, on the scale $\log N$(H)$ = 12.0$, or $Y \approx
0.26$).  In stellar models that take gravitational settling into account, the
surface value of $Y$ can fall to quite small values when the envelope convection
zones become very thin (see V12).  Such variations will affect the mean
molecular weight in the atmospheric layers and presumably have some
repercussions for the temperature structure.

\begin{figure}
\plotone{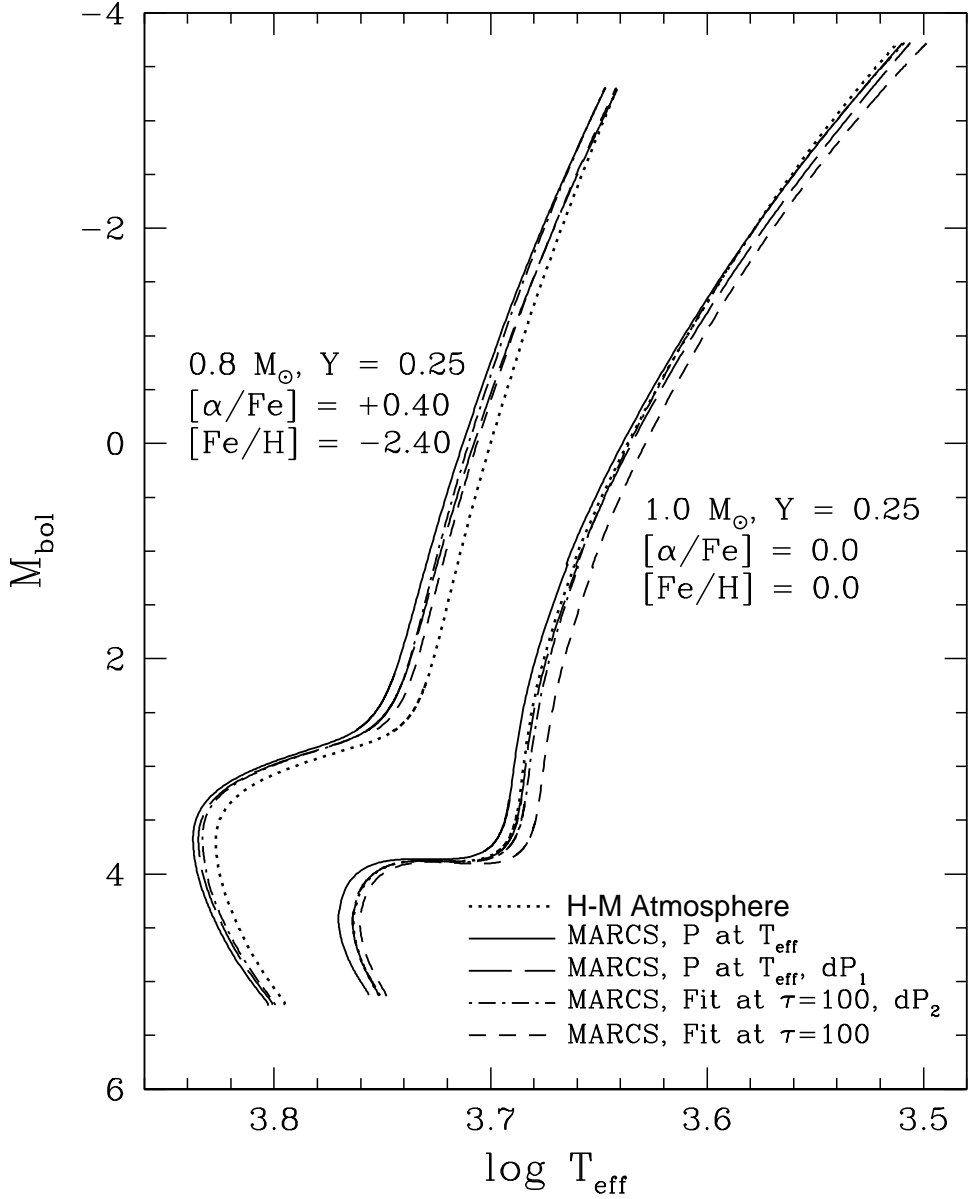}
\caption{Evolutionary tracks for the indicated masses and chemical abundances,
but for different treatments of the atmospheric boundary condition (see the
text).  As noted in the legend in the lower right-hand corner, the dotted curves
indicate the tracks in which a scaled \citet{hm74} $T$--$\tau$ structure was
assumed in determining the surface pressure, while the other loci were obtained
as the result of attaching MARCS model atmospheres (\citealt{gee08}) to the
interior structures at the photosphere or at depth ($\tau = 100$).  In two of 
the latter cases, {\it ad hoc} corrections were applied to the predicted
pressure at $T=\teff$ (dP$_1 = \delta\log P = -0.130$) or at $\tau = 100$
(dP$_2 = \delta\log P = +0.042$) in order for a $1 \msol$, solar abundance model
to reproduce the luminosity and temperature of the Sun at the solar age.}
\label{fig:fig4}
\end{figure}

Even though suitable model atmospheres for use as BCs in the computation of
{\it diffusive} stellar models are not currently available, we did some limited
explorations of the effects on the predicted $\teff$\ scale of using the 
current MARCS grids in this way.  The results of those experiments are shown
in Figure~\ref{fig:fig4}, which plots evolutionary tracks for the indicated
masses and chemical abundances, on the assumption of different treatments of
the atmosphere.  The dotted track for [Fe/H] $=0.0$ differs only slightly
from one that passes through $\log \teff = 3.7617$ ($\teff = 5777$~K) and
$M_{\rm bol} = 4.75$ (the solar temperature and bolometric luminosity) at the
solar age (4.57 Gyr) insofar as it assumed $Y=0.25$ whereas a Standard Solar
Model requires $Y = 0.2553$.  The solid
and dashed loci for the same metallicity assume $\amlt = 2.007$, as in the case
of the dotted curve (and indeed, all other evolutionary sequences that have been
computed), but instead of employing BCs based on the HM74 $T$--$\tau$ relation, 
interpolations in the tabulated properties of the MARCS model atmospheres at
$T = \teff$ or at $\tau = 100$, respectively, were carried out to fit the  
atmospheres to the interior structures.  

In contrast with the findings of \citet{vee08}, who found fairly small
differences between the tracks (for similar masses and chemical compositions)
when MARCS atmospheres were fitted to {\it non-diffusive} stellar models at the
photosphere or at depth, the solid and dashed tracks are appreciably offset from
each other.  Because they run roughly parallel to one another, it seems more
likely that differences in the low-$T$ opacities or of the assumed value of
$\amlt$ is responsible for this separation since, at the zero-age MS location,
diffusion has not had enough time to significantly alter the surface abundances.
Regardless, one could apply suitable adjustments to the pressures predicted by
the MARCS atmospheres, either at the photosphere or at $\tau = 100$, to force
consistency of these cases with the solar constraint.  Doing so results in the
long-dashed and dot-dashed curves, which are not very different from the dotted
track.  The solar calibration thus compensates for most of the differences in
the assumed physics.

However, the computations for [Fe/H] = $-2.40$ show that there are systematic
variations in the tracks as a function of metallicity.  When MARCS atmospheres
are used as BCs, with or without {\it ad hoc} adjustments to the pressures at
the photosphere or at $\tau = 100$, the resultant tracks are all hotter than the
dotted track (by as much as 200 K, see Fig.~\ref{fig:fig4}).  Whether or not
this is telling us that the isochrones presented in this study for low [Fe/H]
values are too cool is hard to say (though it is tempting to conclude that this
is probably the case).  On the one hand, the use of the same $T$--$\tau$
structure for all stellar models, regardless of mass, metallicity, and
evolutionary state, can hardly be realistic.  On the other hand, our isochrones
appear to be able to reproduce the properties of local subdwarfs with
{\it Hipparcos} parallaxes quite well (see \S~\ref{subsec:sbd}, and VandenBerg
et al.~2010).  It is always possible, for instance, that errors associated
with the surface BCs are compensating for those arising from the treatment of
convection or from other physics ingredients.  The observed $\teff$\ scale is
simply not yet precise enough to provide good constraints on the temperatures
predicted by stellar models. 

The final point worth making here is that, as shown by V12, the Victoria and
recent MESA (\citealt{pbd11}) evolutionary codes produce nearly identical tracks
when very close to the same physics is assumed.  Indeed, predictions of such
quantities as the age, luminosity, and helium core mass at the RGB tip are also
in excellent agreement.  The same can be said of the tracks produced by
the Dartmouth code (\citealt{dcj07}), as those computations also appear to be
nearly indistinguishable from ours when the same mass and chemical abundances,
and very similar input physics, are adopted (see \citealt[their Fig.~3]{bvb12}).
Judging from e.g., the plots provided by \citet{bmg12}, such good consistency
between the results of the various evolutionary codes currently in use is not
always obtained, though the extent to which differences in the physics are
responsible for this is not clear. Efforts should be made to understand the
origin of any such discrepancies that are found.

\section{The Adopted Metal and Helium Abundances}
\label{sec:abund}

Although V12 computed numerous grids of models for wide ranges in [Fe/H] that
allowed for variations in $Y$ and [$\alpha$/Fe], as well as for different
heavy-element mixtures in which the abundances of $\gta 10$ metals were varied
in turn, the decision was made to recompute most of them.  Doing so enables us
to adopt the updated solar abundances by \citet{ags09} instead of the
preliminary determinations given by \citet{ags05} (which were assumed by V12),
and to consider (in subsequent papers in this series) $\pm 0.2$ dex variations
in the abundances of those metals which have especially important consequences
for observed CMDs (notably O, Ne, Mg, and Si).

In their exploratory study, V12
investigated the effects of 0.4 dex enhancements in the abundances of individual
metals.  Such variations are too large: in the case of the most abundant
$\alpha$\ elements, star-to-star variations about some representative
[$\alpha$/Fe] value (e.g., 0.4 dex, if [Fe/H] $\lta -1.0$) are typically
$\lta \pm 0.1$--0.2 dex (\citealt{cbg09b}).  In addition, it is of some interest
to determine how both overabundances and underabundances of the various metals
affect computed tracks and isochrones as a function of [Fe/H] given that such
effects are unlikely to have a strictly linear dependence.  (Because of the
overwhelming importance of oxygen for turnoff luminosity versus age relations
and the likelihood that the most metal-deficient stars have [O/Fe] $> 0.4$ 
(\citealt{fna09}; \citealt{rmc12}), the next paper will provide
low-metallicity models in which [O/Fe] varies from 0.2 to 1.0, in 0.2 dex
increments, on the assumption of [$m$/Fe] $= 0.4$ for the other
$\alpha$\ elements.  Sets of models for [$\alpha$/Fe] $= 0.0$, but with
[O/Fe] $= \pm 0.2$ dex, will also be provided.)

The models provided in this paper thus represent the ``base grids" that will
be intercompared with those to be presented in subsequent studies that allow
for variations in the abundances of individual metals.  Here, the same [$m$/Fe]
values ($-0.4$, 0.0, and $+0.4$) have been adopted, in turn, for each of the
$\alpha$ elements (O, Ne, Mg, Si, S, Ar, Ca, and Ti) assuming the $\log N$
abundances for the solar mixture given by \citet{ags09}.  These determinations
are listed in Table~\ref{tab:tab1}, together with the abundances of the
other 11 metals that are considered when OPAL opacities (\citealt{ir96}) are
requested for stellar interior conditions using the Livermore Laboratory web
site.\footnote{http://opalopacity.llnl.gov}  (As in V12, OPAL opacities for
each of the assumed chemical mixtures have been employed, along with the
complementary low-$T$ opacities that have been generated specifically for this
project using the code described by \citealt{faa05}.)  Once the abundances of
the $\alpha$\ elements had been set (to be consistent with the desired value
of [$\alpha$/Fe]), the abundances of all 19 metals were scaled to the [Fe/H]
values of interest simply by adding the latter to the resultant $\log N$ values.
Thus, for the case in which [$\alpha$/Fe] $= +0.4$ and [Fe/H] $= -1.0$, the
$\log N$ abundances of, e.g., C, O, and Fe are 7.43, 8.09, and 6.50,
respectively.  This ensures that, for this particular example, [C/Fe] $= 0.0$
and [O/Fe] $= +0.4$ for the assumed [Fe/H] value, and hence that [C/H] $=$
[Fe/H] $= -1.0$ while [O/H] $ = -0.6$.

Since theoreticians express metal abundances in terms of mass-fractions, $X_i$
(for the $i^{\rm th}$ element), and the total metallicity by the quantity $Z$
--- as opposed to the use of $m$/H number-abundance ratios by observers --- it
is necessary to transform between the different ways of specifying chemical
abundances in order to ensure that stellar models are computed for the {\it
observed} abundances.  Fortunately, both groups of astronomers are comfortable
using $Y$ to describe the helium abundance, but this does cause a slight
complication when converting from number- to mass-fraction abundances (or to 
[$m$/H] values).  Assuming $\log N = 12.0$ for hydrogen and, say, 10.9 (as a
first approximation) for helium, the number-fraction abundances for the $i^{\rm
th}$ element are given by $\zeta_i = N_i/\sum N_i$, where the summation includes
all of the elements that are considered.  If $A_i$ is used to represent the
atomic weight of element $i$, then the corresponding mass-fraction abundances
are given by $X_i = (\zeta_i\,A_i)/\sum (\zeta_i\,A_i)$, and $Z = 1 - X_{\rm H}
- X_{\rm He}$.  By iterating on the value of $\log N$(He) using, e.g., the
secant method, these calculations can be repeated until  $X_{\rm He}$ is equal
to the desired value of $Y$.

From the resultant determinations of $\log N$, the values of [$m$/H] can be
computed using the adopted solar abundances (the second column in
Table~\ref{tab:tab1}).   Note that, if stable isotopes of a given element are
not treated separately, as in the calculation of the number-fraction abundances
that must be specified when generating OPAL opacities, the appropriate
number-weighted value of $A_i$ should be used.  For instance, $A$(C)$ = 12.011$
if $N$($^{12}$C)$/N$($^{13}$C)$ = 90$ and the atomic weights of $^{12}$C and
$^{13}$C are 12.00000 and 13.00336 (\citealt{wb77}), respectively.  By following
these procedures, it is obviously quite easy to obtain essentially exact
equivalences between the different ways of specifying the abundances of the
chemical elements in stars and in stellar models.  In particular, very precise
values of $X_i$ and $Z$ can be determined that correspond to arbitrary values
of [Fe/H] (or, more generally, [$m$/H]).  Put another way: starting with the
tabulated $\log N$ and desired [Fe/H] values, the procedures described above
will yield the correct mass-fraction abundances that should be assumed in the
models which are used to interpret data for the observed metallicities.
 
\section{The Evolutionary Tracks and Isochrones}
\label{sec:vrmod}

\begin{figure}
\plotone{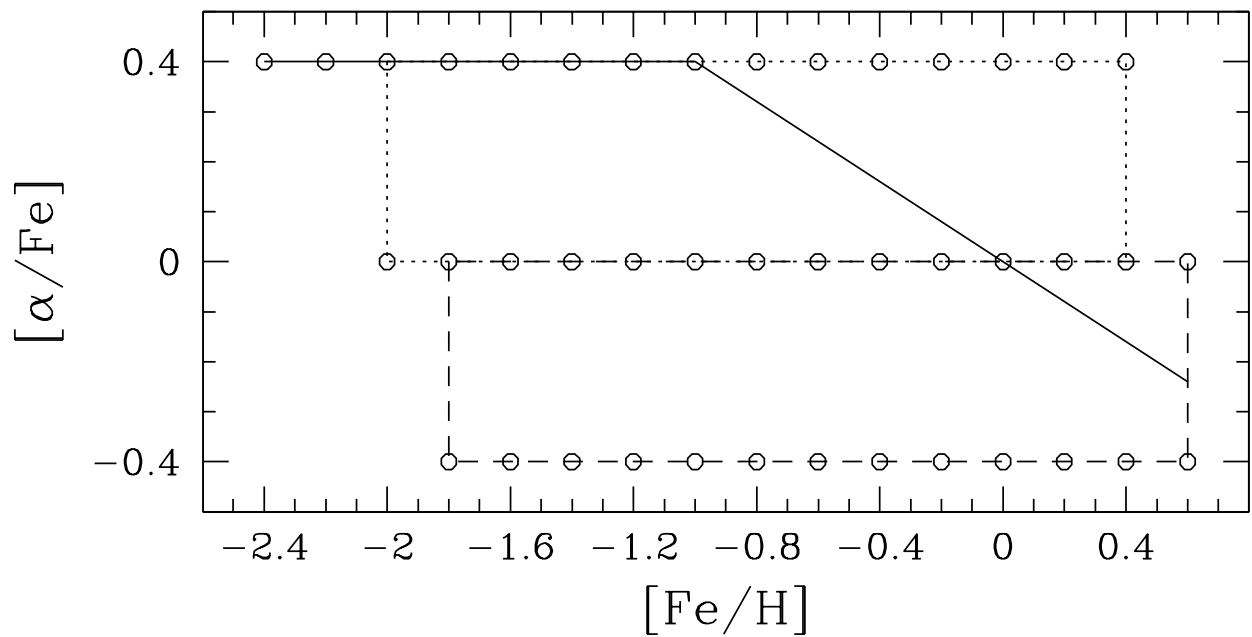}
\caption{Grids of evolutionary tracks for $Y = 0.25, 0.29,$ and 0.33 have been
computed for the values of [Fe/H] and [$\alpha$/Fe] which are designated by the
open circles.  The boxes defined by the dotted and dashed lines indicate the
regions of parameter space that are used when interpolating for models that
have $0.0 \le$ [$\alpha$/Fe] $\le 0.4$ and $-0.4 \le$ [$\alpha$/Fe] $< 0.0$,
respectively (see the Appendix).  The solid curve represents one of many
possible relations between [$\alpha$/Fe] and [Fe/H] that may be assumed when 
modelling a particular stellar population.}
\label{fig:fig5}
\end{figure}

Figure~\ref{fig:fig5} provides a pictorial summary of the values of
[$\alpha$/Fe] and [Fe/H] for which grids of evolutionary tracks have been
computed.  At each of the 42 points which are represented by open circles,
model sequences have been generated for $Y = 0.25$, 0.29, and 0.33, and for
masses that vary from $0.12 \msol$ to a sufficiently high value (ranging between
1.1 and $1.5 \msol$, depending on the metallicity) that its RGB tip age is
$\lta 3$--5 Gyr.  This ensures that isochrones with complete giant branches can
be computed for older ages ($\gta 5$ Gyr).  In general, the tracks were
terminated at the onset of the helium flash, once the He-burning luminosity due
to the triple-$\alpha$ process exceeded $100 L_\odot$, or when the age of the
model reached 30 Gyr, whichever occurred first.  The methods described by
V12, with the recent updates to them that are reported in the Appendix of the
present paper, were used to determine the so-called ``equivalent evolutionary
phase" (EEP) points along the tracks.  It is these EEP files that are
interpolated to produce isochrones for arbitrary values of age, [$\alpha$/Fe],
[Fe/H], and $Y$ (within the ranges encompassed by the model grids; see the
Appendix).

\begin{figure}
\plotone{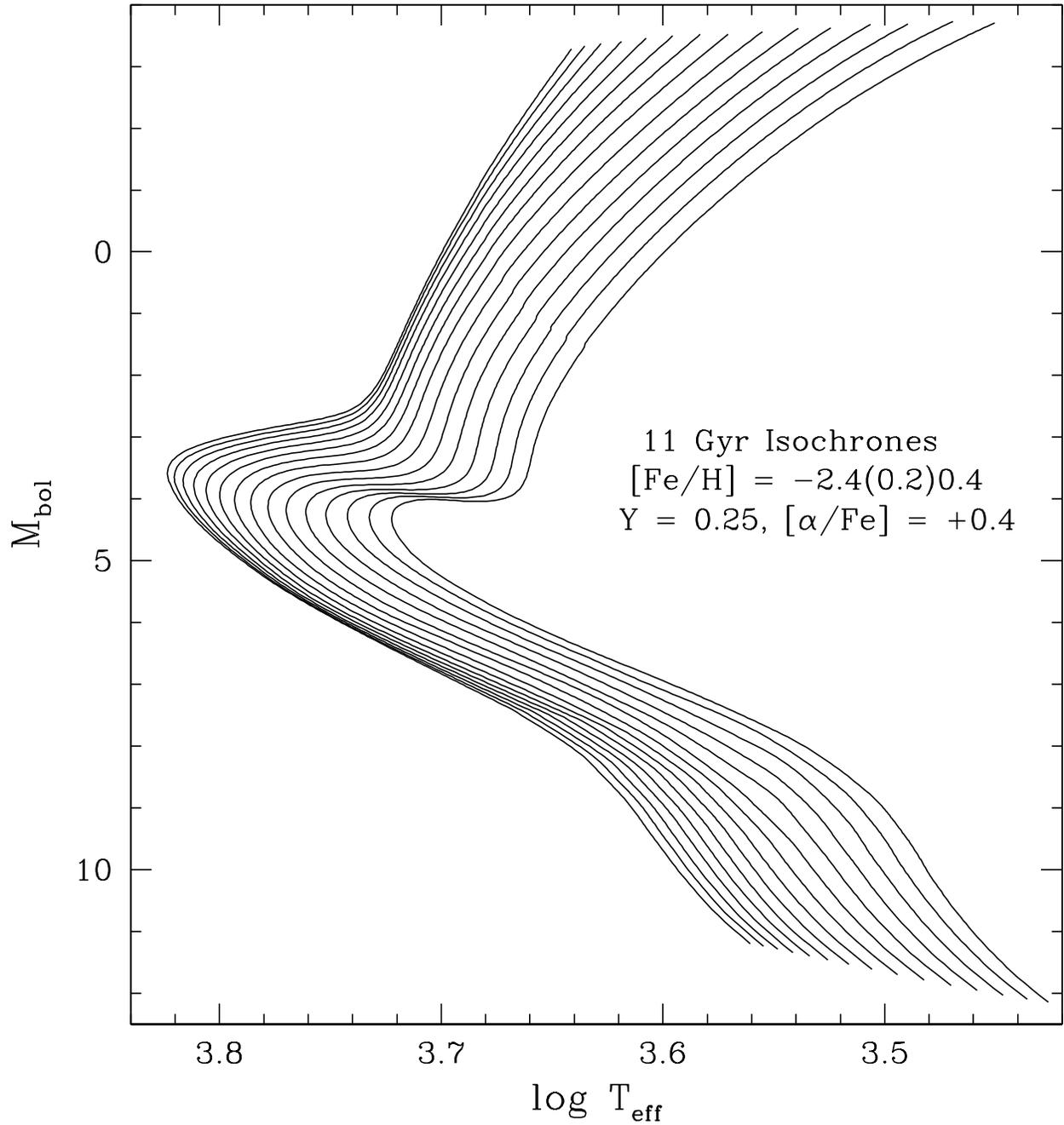}
\caption{Plot on an H-R diagram of 11 Gyr isochrones for the indicated helium
and $\alpha$-element abundances, on the assumption of [Fe/H] values that vary
from $-2.4$ to $+0.4$ (in the direction from left to right), in 0.2 dex
increments.}
\label{fig:fig6}
\end{figure}

The [Fe/H] dependence of isochrones for the same age (11 Gyr), helium abundance
($Y = 0.25$), and value of [$\alpha$/Fe] ($+0.4$) is illustrated in
Figure~\ref{fig:fig6}.  Particularly noticeable are the variations with [Fe/H]
of the slope of the upper RGB and the flattening of the SGB.  The latter is
indicative of the increase in the mass at a given TO luminosity that occurs
as the metallicity, and the opacities in stellar interiors, increase.  Although
not shown, similar plots were prepared for all of the different choices of the
chemical abundance parameters and for a few ages between 5 and 15 Gyr to ensure
that all of the isochrones are well behaved.  (Either linear or spline
interpolations may be employed to derive isochrones from a given evolutionary
track EEP file.  Spline interpolations were used to generate the results that
are shown in Fig.~\ref{fig:fig6}.  Had we opted to use linear interpolations,
the isochrones would not have been quite as smooth, but it would take a close
inspection of the respective plots to identify the very minor differences.) 

Because we use the very efficient non-Lagrangian method devised by
\citet{egg71} to follow RGB evolution (see \citealt[2012]{van92}), the location
of the so-called ``RGB bump" is much less obvious, if at all, in our tracks than
in those generated using a Lagrangian code, because the Eggleton technique does
some numerical smoothing of what is predicted to be a very sharp boundary when
mass is taken to be the independent variable.  (When the H-burning shell passes 
through the chemical abundance discontinuity that was produced near the base of
the giant branch by the deepest penetration of the convective envelope,
Lagrangian models will generally evolve to slightly lower luminosities as the
stellar structure adjusts to a somewhat higher hydrogen abundance before
continuing up the RGB.  The additional time spent in the small luminosity range
where this occurs manifests itself as a local enhancement in the differential
luminosity function, which is commonly referred to as the RGB bump.) In our
models, the evolution stalls during this adjustment phase and the bump
luminosity is easily identified (even at metallicities as low as [Fe/H]
$= -4.0$) as a local minimum or maximum in $d(\log\,L)/d(\log\,t)$ or
$d(\log\,\teff/d(log t)$, respectively, where $t$ represents time (e.g.,
\citealt[their Fig.~2a]{bv92}).  However, as shown by, e.g., \citet{pcs07}, the
luminosity functions derived from our isochrones, including the locations of
the RGB bump, agree very well with similar results from Lagrangian codes. 

\begin{figure}
\plotone{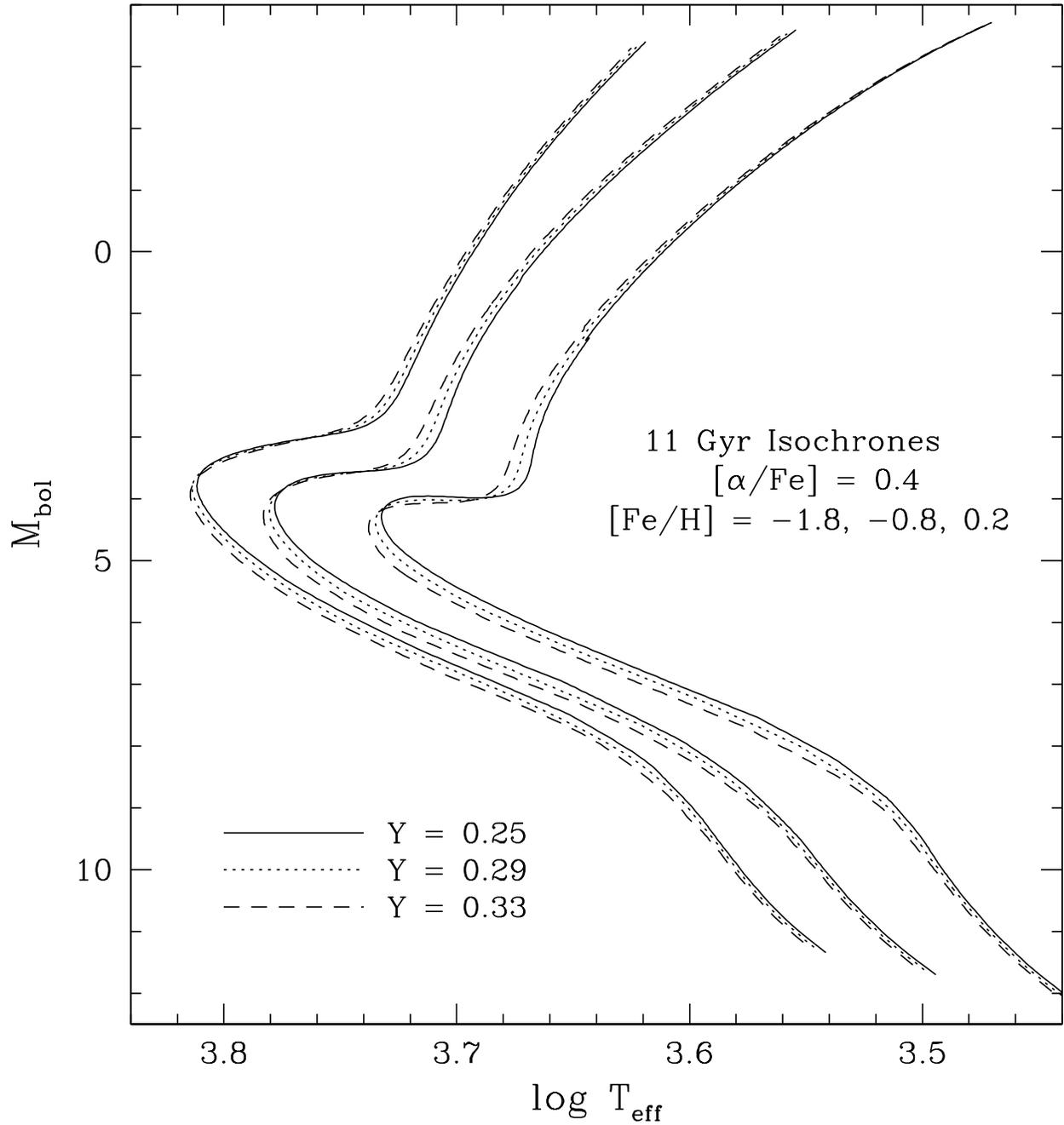}
\caption{Comparison of three of the isochrones from the previous figure (i.e.,
for $Y = 0.25$, [$\alpha$/Fe] $=0.4$ and [Fe/H] $= -2.0$, $-1.0$, and 0.0:
solid curves) with isochrones which are otherwise identical, except that
the assumed helium abundances are $Y = 0.29$ (dotted curves) and
$Y = 0.33$ (dashed curves).}
\label{fig:fig7}
\end{figure}

Figures~\ref{fig:fig7} and~\ref{fig:fig8} illustrate, in turn, the effects on
11 Gyr isochrones for [Fe/H] $= -1.8$, $-0.8$, and $+0.2$ of varying the helium
and [$\alpha$/Fe] abundances. (Very similar plots have been provided by
\citealt{vcs12}.) Interestingly, helium apparently has bigger
consequences for the temperatures of faint giants than those near the RGB tip,
and the morphology of the SGB is more sensitive to $Y$ at high [Fe/H] values
than it is at low metallicities.  Opposite to the effects of the metals on
mass-luminosity relations (as noted above), an increased helium abundance
implies a lower mass at a fixed TO luminosity due primarily to the concomitant 
change in the mean molecular weight throughout the structure of a star.  At
intermediate and high [Fe/H] values, isochrones are affected more by variations
in the abundances of the $\alpha$-elements than of helium: compare 
Figs.~\ref{fig:fig7} and~\ref{fig:fig8}, which also shows that the ramifications
of increasing [$\alpha$/Fe] from $-0.4$ to 0.0 tend to be smaller than those
associated with an increase from 0.0 to $+0.4$. (It is worth mentioning
that photometry may be used to constrain the value of [$\alpha$/Fe] in
relatively simple stellar populations since both the location and slope of the
upper RGB and, in particular, the LMS portions of isochrones are predicted to
be quite sensitive to [$\alpha$/Fe] on some CMDs; see \citealt[their
Fig.~16]{cv14}.)  Note that the variations in the TO (and SGB) luminosities at
a fixed age and [Fe/H] are mainly due to the different oxygen abundances in the
three [$\alpha$/Fe] cases (see V12). 

A few additional remarks are warranted concerning the striking difference 
between the dotted and dashed isochrones for [Fe/H] $= +0.2$ given that the
separation betweeen their LMS segments becomes quite large at $M_{\rm bol}
\gta 9$, in contrast with their behavior at lower metallicities.  In fact, the
upper giant branches of the same isochrones also seem odd in that they merge
near the tip, rather than running approximately parallel to each other, as in
the case of the RGBs for [Fe/H] $= 0.0$ and $+0.4$.  To investigate the cause(s)
of such differences, we compared the tabulated low-temperature opacities for the
three values of [$\alpha$/Fe] at fixed values of temperature, density, and
[Fe/H], and made the unexpected discovery that the opacity variations between
the tables for [$\alpha$/Fe] $= +0.4$ and 0.0 were very different from those
between the tables for [$\alpha$/Fe] $= 0.0$ and $-0.4$.  We then realized that
there is a fundamental difference in the three mixtures; namely, that the
C/O ratio is $< 1$ if [$\alpha$/Fe] $= -0.4$, but that it is $> 1$ in the case
of the two higher values of [$\alpha$/Fe] (see Table~\ref{tab:tab1}).  This can
have a huge impact on the opacity at $T \lta 3.42$ (see \citealt[their
Fig.~4]{fd08}), and thereby on the surface boundary conditions and predicted
effective temperatures of stellar models that have sufficiently cool outer
atmospheric layers.  This undoubtedly explains the seemingly anomalous behavior
of the LMS and upper RGB portions of the [Fe/H] $= +0.2$ isochrone for
[$\alpha$/Fe] $= -0.4$ relative to those for the same [Fe/H] and [$\alpha$/Fe]
$= 0.0$ and $+0.4$.

(In view of these findings, we plan to do a more thorough investigation of
models for [$\alpha$/Fe] $= -0.2$ to $-0.4$ at a later date.  Because the
opacity at $\log\,T \lta 3.42$ changes very rapidly as the C/O ratio varies
from $> 1$ to $< 1$, the LMS and upper RGB portions of the interpolated
isochrones for [Fe/H] $\gta -0.4$ and sub-solar abundances of the
$\alpha$-elements {\it may} be discrepant relative to (non-interpolated)
isochrones based on stellar models that have been computed for those abundances.
Whether or not the discrepancies are significant will not be known until the
planned grids for a finer spacing in [$\alpha$/Fe] have been computed.  This is
not a concern for interpolations in our current grids for [$\alpha$/Fe]
$\ge 0.0$ at any [Fe/H] value or in those for [$\alpha$/Fe] $< 0.0$ if [Fe/H]
$\lta -0.4$.  Even at higher metallicities, the interpolated isochrones for
[$\alpha$/Fe] $< 0.0$ will be fine for the upper MS, TO, and lower RGB phases.)

\section{Selected Comparisons of the Models with Observations}
\label{sec:obs}

Although star cluster CMDs are often used to test stellar models (e.g.,
\citealt{dvd12}, \citealt{bmg12}), and indeed, they do provide valuable
constraints on such aspects of stellar physics as the extent of convective core
overshooting (e.g., \citealt{mrr04}; \citealt{vbd06}) and on the variation of
the mixing-length parameter, $\amlt$, with metallicity (e.g., \citealt{fvs06}),
it should be kept in mind that they are also the targets of astrophysical
reasearch and that their basic properties (distances, reddenings, chemical
abundances) involve appreciable uncertainties.  For instance, estimates
of the iron content of the GC M$\,$15 range from values as high as [Fe/H] $\sim
-2.1$ (\citealt{cg97})\footnote{These investigators currently appear to favor a
somewhat lower value (specifically, $-2.33$, see \citealt{cbg09a}), but this
value of [Fe/H] is based, in part, on the adoption of a cooler $\teff$\ scale
than in the 1997 study.  Had hotter temperatures been assumed, they would
(presumably) have derived higher values of [Fe/H] from the observed spectral
line strengths.} to as low as $\sim -2.6$ (\citealt{pst06}, \citealt{sks11}).
Having such a wide range in the measured [Fe/H] values compounds the difficulty
of evaluating predictions for, e.g., the location of the giant branch relative
to the turnoff, which is known to depend quite strongly on both age and metal
abundance (see, e.g., \citealt[their Figs.~1--3]{vbl13}), as well as the
treatment of convection and the atmospheric boundary condition (among other
things).  Because of the many uncertainties at play, including those associated
with color-$\teff$ relations, isochrones cannot be expected
to provide perfect matches to observed CMDs.  For the same reason, any
discrepancies that occur are not easy to explain --- and this situation will
likely continue until the observed metallicity, temperature, and distance
scales are much better determined than they are at the present time.

Field subdwarfs with accurate parallax-based distances have long been recognized
as important Population II standard candles (along with RR Lyrae variables and
white dwarfs), but their usefulness is also limited by chemical composition and
temperature uncertainties.  Fortunately, high resolution, high signal-to-noise
spectra are readily obtained for them, due to their proximity, but analyses of
spectroscopic data (in general) are complicated by non-LTE and 3D effects, which
appear to be particularly important for metal-poor stars (\citealt[and
references therein]{mca13}).  Consequently, significant revisions to the basic
properties of local subdwarf stars may well occur in the coming years as more
and more sophisticated model atmospheres are employed in analyses of their
spectra.  In any case, with these few cautionary remarks, we will now present
and discuss a few comparisons of our models with observations.  In fitting
isochrones to photometric data, we have employed the color--$\teff$\ relations
derived by \citet{cv14} from the latest MARCS model atmospheres
(\citealt{gee08}).

\subsection{The Solar-Metallicity Open Cluster M$\,$67}
\label{subsec:m67}

M$\,$67 is younger than the age range focused on by the present investigation,
but it is still worthwhile to fit isochrones to its CMD in order to see how
well solar abundance models reproduce its MS and RGB.  High-resolution
spectroscopy has revealed that this system has nearly the same [Fe/H] value as
the Sun and very close to the solar mix of the metals (\citealt{rsp06},
\citealt{okg11}).\footnote{The most recent determination of the iron content of
M$\,$67 is by \citet{ogk14}, who obtained [Fe/H] $= +0.06$.  This includes a
correction to the observed abundances (based on theoretical models) to take
the effects of metals diffusion into account.}  Moreover, as both the earlier
dust maps by \citet{sfd98} and the recent recalibration of them by
\citet[hereafter SF11]{sf11} yield a reddening that is consistent with $E(B-V)
= 0.030 \pm 0.003$, the Sun may be used to provide quite an accurate estimate
of the cluster distance modulus via the MS-fitting technique because of the
similarity in their ages: according to most estimates, M$\,$67 is only $\sim
0.4$--0.7 Gyr younger than the Sun (see, e.g., \citealt{rfr98}, \citealt{mrr04},
\citealt{okg11}).
 
As predictions of synthetic magnitudes for longer wavelength filters are likely
to be less problematic than those for blue or ultraviolet bandpasses, we have
opted to fit isochrones to the $V\,K_S$ photometry for M$\,$67 that was analyzed
by \citet{bsv10}.  The latter combined a recent reduction of Johnson-Cousins
$V$-band data for M$\,$67 (by one of the co-authors; namely, P.~B.~Stetson, see
\citealt{st05}) with 2MASS $K_S$ photometry (\citealt{scs06}) to produce a CMD
that is well defined down to $V \sim 19$.  After the colors of the stars have
been dereddened, assuming $E(V-K_S) = 2.76\,E(B-V)$ (\citealt{cv14}), and their
apparent magnitudes have been decreased by 9.69 mag (the adopted apparent
distance modulus), one obtains the CMD that is shown in Figure~\ref{fig:fig9}.
The vertical offset was chosen so that the cluster MS lies just slightly fainter
than the Sun (which has $M_V = 4.82$) at the solar color ($V-K_S = 1.560$;
\citealt{crm12}), to be be consistent with the aforementioned age difference.
The isochrone that provides the best fit to the cluster subgiants has an age of
4.3 Gyr. [Interestingly, the age corresponding to a given magnitude difference
between the SGB and the MS, at a fixed color, depends quite sensitively on the
adopted value of $Z$.  As shown by \citet[their Fig.~2]{vge07} and
\citet{mrr04}, ages $\lta 4$ Gyr are obtained for M$\,$67 if $Z \gta 0.017$,
and vice versa.]  Encouragingly, the MARCS transformations yield the same
$V-K_S$ color for the Sun as that derived by Casagrande et al.~to within
0.006 mag.

In their study of M$\,$67, \citet{bsv10} employed the 3.7 Gyr isochrone that 
was derived by \citet{mrr04} from models that took gravitational settling and
radiative accelerations into account (and they made somewhat different 
assumptions about the solar normalization and initial abundances).  Unlike our
isochrone, it provides an excellent fit to the morphology of the turnoff,
including the luminosity of the gap.  In order for our computations to have
similar success, it will be necessary to consider diffusion of the metals
and to calibrate the extent of convective core overshooting.  However, our
interest here is not to derive the age of M$\,$67, but rather to ascertain how
well our solar abundance models are able to reproduce those parts of a CMD
(namely, the MS and RGB) that have no more than a weak dependence on age, and
in addition, are virtually independent of whether or not diffusive processes and
core overshooting are treated.  Indeed, in these respects, our models fare quite
well as they provide a very good match to both the cluster giant branch and
the lower MS (down to $M_V \sim 8.5$).  This suggests that the predicted $\teff$
scale for solar metallicity stars is accurate over a wide range in luminosity.
These results also provide support for stellar models (at least those for [Fe/H]
$\approx 0.0$) that use the scaled HM74 $T$--$\tau$ relation to derive the
pressure at $T=\teff$.

\subsection{The Super-Metal-Rich Open Cluster NGC$\,$6791}
\label{subsec:n6791}

NGC$\,$6791 is one of the oldest and most metal-rich open clusters known, and
for these reasons it has been the subject of numerous investigations over the
years (e.g., \citealt{gvz94}, \citealt{ovr06}, \citealt{bbg11}).  Because its
CMD is characterized by tight, well-defined photometric sequences (e.g.,
\citealt{sbg03}), and because two of its eclipsing binaries have been subjected
to careful analyses (see \citealt[2011]{bvb12}), NGC$\,$6791 provides an
especially powerful probe of the properties of old, super-metal-rich stars.  In
particular, the mass-radius (MR) diagram for the observed binaries provides an
important constraint on the helium content of NGC$\,$6791 if the abundances of
the metals are obtained from spectroscopic data.  Moreover, an estimate of the
binary, and hence cluster, distance that is completely independent of CMD
considerations may be derived from the luminosities which are implied by their
radii and, say, spectroscopically determined values of $\teff$.  \citet{bvb12}
concluded that NGC$\,$6791 has an age of $\approx 8.3$ Gyr if it has [Fe/H]
$= +0.35$ (with the metals in the proportions given by \citealt{gs98}),
$Y = 0.30$, $E(B-V) = 0.14$, and $(m-M)_V = 13.51$.

According to K.~Brogaard (2013, priv.~comm.), a slightly lower [Fe/H] value 
($\lta 0.30$) seems to be favored by the latest spectroscopic results (in
agreement with the earlier findings of \citealt{bjd09}).  If [Fe/H] $= 0.30$ is
adopted, it is a straightforward and relatively quick exercise to iterate
between the fits of isochrones for different $Y$ and age to the MR
diagram for the binaries known as V18 and V20 (for numerical values of their
properties, see \citealt[their Table 1]{bvb12}) and the cluster CMD to obtain
the best possible consistency between them.  We have opted to use the $V\,J$
photometry for NGC$\,$6791 compiled by \citet{bsv10}: they collected new $J$
observations, which were calibrated to the 2MASS system (\citealt{scs06}) and
then combined with $V$-band data for the same stars (e.g., \citealt{st05}).

Initially, we found that an 8.5 Gyr isochrone for [Fe/H] $= 0.30$ and $Y =
0.28$ provided quite a good fit to the photometry if $E(B-V) = 0.14$, which
agrees well with the value of 0.133 from the SF11 dust maps, and $(m-M)_V =
13.55$.  However, in a separate (concurrent) study, \citet{cv14} were able to
obtain a consistent fit of the same isochrone to most of the CMDs that can be
constructed for NGC\,6791 from publicly available $BVIJ$ and Sloan $ugriz$
photometry if $E(B-V) = 0.16$ and the equivalent true distance modulus,
$(m-M)_0 = 13.05$, are assumed.\footnote{In the paper by \citet{cv14},
reddening is treated in a fully self-consistent way; i.e., the dependence of
the color excess on the spectral type of a star is correctly taken into
account using tables of reddening-corrected bolometric corrections (BCs).  Thus,
e.g., a value of $E(B-V)$ that is appropriate for early-type stars, which is
the usual convention for reddenings reported in the literature,
would be less for a turnoff star in NGC\,6791 by about 10\%.  Color excess
ratios such as $E(V-I)/E(B-V)$ also vary with spectral type.  (If the 
reddening is low, it is reasonable to assume that the extinction coefficient
in a given band, $R_\lambda$, is constant, though the adopted values of these
quantities should be approximately correct for the spectral type of the star(s)
under consideration.  The $R_\lambda$ values applicable to turnoff stars that
have $5250 \le \teff \le 7000$~K and $-2.0 \le$ [Fe/H] $\le +0.25$, along with
extensive tables of BCs for $0.0 \le E(B-V) \le 0.72$, are provided by
Casagrande \& VandenBerg for the majority of the broad-band photometric systems
currently in use.)}  Curiously, the most problematic CMD turned out to be the
same [$(V-J),\,V$]-diagram that we have considered here.  If $E(B-V) = 0.16$ is
adopted, as implied by most of the observations considered by Casagrande \&
VandenBerg, the observed colors and/or the transformations to $V-J$ must be
corrected by a combined total of 0.04 mag (which is equivalent to a change of
0.018 mag in $E(B-V)$ since $E(V-J) \approx 2.23\,E(B-V)$; see \citealt{cv14}),
in order to obtain a good fit of the isochrone to the Brasseur et al.~CMD.
[Further work will be needed to identify the cause of the color offset if,
indeed, the foreground reddening in the direction of NGC\,6791 truly is
$E(B-V) = 0.16$.]

Figure~\ref{fig:fig10} shows that the observed RGB is redder (at $V \lta 15.5$)
than the isochrone which otherwise does a fine job of reproducing the fainter
photometry.  This suggests that either the model temperatures along the upper
giant branch are too hot, or the adopted color--$\teff$\ relations for low
gravities yield $V-J$ colors that are too blue, or both.  As \citet{bsv10} did
not obtain $V\,J$ data for fainter MS stars than those plotted in
Fig.~\ref{fig:fig10}, we are unable to comment on how well the isochrone fits
near-IR data for LMS stars.  However, the plots provided by
\citet[see their Figs.~12 and 13]{cv14} indicate that the same isochrone tends
to deviate to the blue of the cluster MS at 2--3 mag below the turnoff
(depending on the selected color index), while providing a comparable fit to
the upper MS, TO, and SGB stars as that shown in Fig.~\ref{fig:fig10}.
Presumably, the discrepancies at faint magnitudes are also indicative of errors
in the model $\teff$ scale and/or the color transformations for cool,
super-metal-rich stars.

The dashed curve in this figure represents an isochrone for 8.0 Gyr, $Y = 0.30$,
and [Fe/H] $=+0.35$ that has been computed assuming the $m$/Fe number abundance
ratios given by \citet{gs98} instead of those determined by \citet{ags09}.  As
mentioned in the first paragraph of this section, \citet{bvb12} derived a 
slightly higher age (8.3 Gyr) on the assumption of exactly the same chemical
abundances.  This is, however, the expected consequence of their adoption of a
smaller value of $(m-M)_V$ by 0.04 mag, which is easily within the $1\,\sigma$
uncertainty associated with the cluster distance modulus (see the Brogaard et
al.~paper for a discussion of this issue).  Fig.~\ref{fig:fig10} shows that,
with just a small difference in age, and minor changes to the adopted values of
[Fe/H] and $Y$, isochrones based on either of the \citet{gs98} or \citet{ags09}
metals mixtures provide equally good fits to the CMD of NGC\,6791 (as well as
its binaries, see below).  This reinforces the conclusions reached by Brogaard
et al.~from a similar analysis that such comparisons between theory and
observations are not able to provide a clear preference for either solar
abundance mixture, due in part to the compensating effects of the respective
solar calibration.

The masses and radii that were determined for the components of the binaries
V18 and V20 in NGC$\,$6791 by \citet{bvb12} are shown in Figure~\ref{fig:fig11},
along with the predicted MR relations from four different isochrones that
assume scaled \citet{ags09} metal abundances and one isochrone which assumes the
same $m$/Fe ratios that were derived for the Sun by \citet{gs98}.  The 
{\it solid} and {\it dashed} curves represent the same isochrones that were
plotted in the previous figure, and both provide reasonably good fits to the
data.  Indeed, very similar plots are given by Brogaard et al., who showed that
it is only when $3\,\sigma$ error boxes are plotted that the observations can
be intersected by a single isochrone.  At this stage, it is not known whether
the apparent discrepancies are due more to deficiencies in the models that have
been compared with the observed masses and radii or to errors in the derived
properties of the binaries.  It would certainly be worthwhile to collect and
analyze more observations of them and to add to the sample of completely
eclipsing binaries that have been discovered to date in NGC$\,$6791.

Be that as it may, Fig.~\ref{fig:fig11} shows how the MR relation that is
represented by the solid curve would be altered by, in turn, a 0.5 Gyr increase
in age (the dot-dashed curve), a 0.05 dex reduction in [Fe/H] (the dotted
curve), and a change in $Y$ by $+0.01$ (the long-dashed curve). According
to these results, we would have obtained a closer match of the dashed locus to
the solid curve, with an equally good fit to the cluster CMD, if the former
assumed [Fe/H] $= +0.34$ (instead of $+0.35$) or a larger helium abundance by
$\delta\,Y \approx 0.002$.  (The effects of variations in [$\alpha$/Fe] have
not been considered because, to within the uncertainties, the abundances of the
$\alpha$ elements appear to be consistent with scaled solar values; see
\citealt{bbg11}.)

To conclude: aside from a possible zero-point error in the near-infrared
photometry that we have used (\citealt{bsv10}) or in the \citet{cv14}
transformations to the $J$-band, our isochrones for [Fe/H] $\sim 0.3$ are able
to reproduce the observed [$(V-J),\,V)$]-diagram of NGC\,6791 quite well in a
systematic sense --- at least at $V-J \lta 2.2$, which corresponds to $\teff
\gta 4600$\ K.  (Especially encouraging comparisons between the same isochrones
and many other CMDs of this open cluster, derived from available $BVI$ and
$ugriz$ photometry, are provided by Casagrande \& VandenBerg.)  We have also
demonstrated that it is easy to use the models presented in this study to
iterate on the age and chemical abundance parameters until a consistent fit is
found to both an observed CMD and the MR relation that can be obtained from
observations of detached, eclipsing binaries that belong to the same cluster.

\subsection{Local Subdwarfs With $-2 \lta$ [Fe/H] $\lta -1$}
\label{subsec:sbd}

VandenBerg et al.~(2010) have already shown that current Victoria-Regina stellar
models satisfy the constraints provided by subdwarfs in the solar neighborhood
that have well-determined $M_V$ values from {\it Hipparcos}.  In fact, good
consistency between theory and observations is obtained on several different
color-magnitude planes, particularly those involving red or near-infrared
colors (also see \citealt{bsv10}), or on the $(\log\teff,\,M_V)$-diagram if the 
temperatures of the Population II dwarfs are obtained from \citet{crm10}.
The $\teff$\ scale derived by the latter is $\sim 150$\ K hotter than the one
by \citet[\citealt{aam96}]{aam99}, which was widely adopted during the last
decade, though it agrees well with the hot temperature scale first proposed by
\citet{ki93}, and three years later by \citet{gcc96}.  It may be recalled that
the [Fe/H] values determined for GCs by \citet{cg97} are based, in part, on a
hot $\teff$\ scale.

It is of some interest to revisit the work by R.~G.~Gratton and collaborators
in the late 1990s, as their $\teff$ and [Fe/H] estimates for local subdwarfs
are in remarkable agreement with the predictions of present-day isochrones.
This is shown in Figure~\ref{fig:fig12}, which plots (in the bottom panel)
the absolute visual magnitudes for the 10 subdwarfs with [Fe/H] $\lta -1.0$
that have the smallest values of $\sigma_\pi/\pi$ (\citealt{vl07}), where $\pi$
represents the trigonometric parallax, as a function of their effective
temperatures.  The sources of the $\teff$ (and [Fe/H]) determinations are
Gratton et al.~(1996), \citet{gfc97}, \citet{cgc99}, and R.~G.~Gratton (2001,
priv.~comm., as reported by \citealt[see their \S~4.1]{bv01}).  If the models
provided a perfect match to the observed stars, each of the subdwarfs would sit
on the isochrone for its measured [Fe/H] value and the temperature implied by
that isochrone would be identical to the spectroscopic estimate of $\teff$.
(Of course, even the best metallicity and $\teff$\ determinations are uncertain
by $\sim \pm 0.2$ dex and $\sim \pm 70$\ K, respectively.  Furthermore, the ages
of the subdwarfs could well be higher or lower than 12 Gyr --- though most of
them are sufficiently faint that the effect of the age uncertainty will have
negligible consequences for our comparisons with the observations.) 

The middle panel plots, as a function of the temperatures of the subdwarfs, the
differences between the [Fe/H] values that were determined spectroscopically
and those inferred from the isochrones that match the subdwarf locations in the
$(\log\teff,\,M_V)$-diagram.  For the sample of 10 stars, the mean offset is only
0.04 dex, in the sense that the observed iron abundances are just slightly less
than the values deduced from the isochrones, with a standard deviation of 0.29
dex.  Interestingly, the differences between the observed (``Obs") and isochrone
(``Iso") metallicities tend to be $\gta 0.0$ for stars that have [Fe/H] $\ge
-1.5$ (those represented by open circles) whereas more metal-deficient
stars, which are plotted as filled circles, all have ``Obs $-$ Iso" values
$< 0.0$.  However, there is no obvious variation of $\delta\,$[Fe/H] with
temperature for either group of stars, which suggests that the models predict
the correct lower-MS slopes.

One may alternatively interpolate in the isochrones to determine how much of an
adjustment to the temperature of each subdwarf, at its observed $M_V$, would be
required to locate it on the isochrone that has the same [Fe/H] as the subdwarf.
The differences in $\teff$\ so derived are plotted in the upper panel of
Figure~\ref{fig:fig12}.  Not surprisingly (because the abundance implied by a
given line strength depends directly on the adopted temperature), stars with
[Fe/H] $\ge -1.5$ tend to have ``Obs $-$ Iso" values of $\delta\,\teff \gta
0.0$, while the opposite is found for the most metal-poor stars.  As in the
middle panel, the level of agreement is surprisingly good: the mean offset and
standard deviation are only $-11$\ K and 65 K, respectively.  Though the sample
of stars is small, the models appear to fit the observations equally well over
the entire temperature range encompassed by the stars.  (Considering just the
10 subdwarfs in our sample, the temperatures and [Fe/H] values determined by
R.~G.~Gratton and collaborators are, in the mean, 17 K and 0.09 dex higher,
respectively, than the values tabulated by \citealt{crm10}.)  

The same isochrones, when plotted on the [$(V-I)_0,\,M_V$]- and
[$(V-K)_0,\,M_V$]-diagrams, provide equally satisfactory fits to the same
subdwarfs --- as shown in Figure~\ref{fig:fig13}.  The differences in the
predicted colors, which are based on the MARCS color--$\teff$ relations
(\citealt{cv14}), are obviously in excellent agreement with those observed (from
\citealt{crm10}) since, on both color planes, $<\delta$(color)$>\ = 0.00$, with
relatively little scatter about the horizontal dashed line.  Moreover, the
models appear to fit the brighter, bluer stars just as well as the reddest,
faintest ones.  It is important to appreciate that relatively high temperatures
{\it must} be assumed for the subdwarfs in order for the MARCS transformations
to yield the observed colors.  Most broad-band colors (especially $V-I$ and
$V-K$) are much more dependent on $\teff$ than on [Fe/H] (or on gravity, which
will, in any case, be close to $\log\,g = 4.5$ for the Population II dwarfs).
That is, our isochrones are able to provide good fits to the observations only
because they predict the particular $\teff$\ scale that yields the observed
colors when derived from the MARCS color--$\teff$ relations.  Similar success
would not have been obtained had the latter predicted much redder or bluer
colors at the same $\teff$ or if we had adopted a significantly cooler empirical
$\teff$ scale (e.g., Alonso et al.~1999, 1996).

Despite the indications from the spectroscopic results described above, the
recent calibration of the infra-red flux method (IRFM) by \citet{crm10}, the
predictions of our stellar evolutionary models, and the color--temperature
relations implied by the latest MARCS model atmospheres in support of a hot
$\teff$ scale, it is important to remember that current 1D model atmospheres
play a central role in each of these avenues of research.  Indeed, as discussed
by Magic et al.~(2013), the very different temperature structures produced by
3D model atmospheres, particularly at low $Z$, are bound to impact $\teff$ and
[$m$/H] determinations, as well as color transformations and the boundary
conditions employed by stellar models.  Indeed, the importance of advancing our
understanding of model atmospheres, which provide the interface between stellar
interior models and observed stars and stellar populations, can hardly be
understated.

\subsection{The Globular Clusters 47 Tuc, M$\,$3, M$\,$5, and M$\,$92}
\label{subsec:gcs}

In their extensive survey of GC ages, \citet{vbl13} found that isochrones
generally provided reasonably good fits to the {\it Hubble Space Telescope}
ACS (Advanced Camera for Surveys) photometry obtained by \citet{sbc07} when
the cluster distances were determined from fits of zero-age horizontal-branch
(ZAHB) loci to the observed HB stars.  All of the models used in that
investigation assumed the solar metal abundances given by \citet{gs98}, with
suitable enhancements to the abundances of $\alpha$-elements and then scaled to
the [Fe/H] values derived by \citet{cbg09a}.  As we have not yet computed ZAHBs
for the chemical mixtures assumed in this study, we are unable to follow exactly
the same procedure here in order to ascertain, in particular, how the inferred
distance moduli will differ from those found by VandenBerg et al.  However, the
predicted ZAHB luminosities, at the same [Fe/H], are likely to be quite similar 
because the main difference in the solar mixtures given by Grevesse \& Sauval
and \citet{ags09} are the abundances of CNO, which mainly affect the color of
the HB.\footnote{Complementary ZAHB loci will be provided in a later paper, once
additional grids of models for the MS, RGB, and HB phases have been computed
that allow for variations in [O/Fe].  Since the majority of low-metallicity
([Fe/H] $\lta -1.0$) stars in the Milky Way appear to have [O/Fe] $\gta 0.6$
(see, e.g., Ram\'irez et al.~2012), it is our intention to provide the means to
interpolate in the resultant grids to obtain ZAHB sequences (and isochrones) 
for different oxygen abundances at the same values of [Fe/H] and $Y$ (assuming
[$m$/Fe] $= 0.0$ and $+0.4$ for the other $\alpha$-elements).  A further
advantage of presenting all of the ZAHB models in the same paper is that our
discussion of them will be considerably simplified.}  

Because TO luminosity versus age relations depend sensitively on the absolute
abundance of oxygen (see V12), and because our models assume a smaller value
of [O/Fe] (by 0.1 dex) as well as a lower solar abundance of oxygen, it can be
expected that we will obtain higher ages for metal-poor clusters than those
derived by \citet{vbl13} (if all other variables are kept constant).   However,
this is a moot point for the present discussion.  Our main motivation for
examining the CMDs of a few GCs is to check how well our models are able to
reproduce the observed MS and RGB morphologies.  To partially compensate for the
expected effects of the different abundances of oxygen noted above (and of other
metals), we have {\it arbitrarily} assumed slightly larger distance moduli (by
$\lta 0.05$ mag) and ages (by 0.25 Gyr) than the values derived by VandenBerg
et al., and then matched the predicted and observed turnoffs.  To accomplish
this, it was necessary to apply a small blueward shift to the isochrones (by
$\lta 0.02$ mag) after the observed colors had been
dereddened.\footnote{\citet{vbn14} have found that such color offsets
would be reduced, if not eliminated entirely, if the GC [Fe/H] scale were
adjusted to lower values by 0.2--0.3 dex to be consistent with the findings
of recent spectroscopic studies of M\,15 (\citealt{pst06}, \citealt{sks11})
and M\,92 (\citealt{rs11}), and by 0.1--0.15 dex at metallicities appropriate
to more metal-rich clusters, such as M\,5.  However, this is just one of many
possible explanations of differences between predicted and observed turnoff
colors (see \citealt[their \S~6.1.2]{vbl13}).}  The result of this exercise
is shown in Figure~\ref{fig:fig14} for the GCs 47 Tuc, M$\,$3, M$\,$5, and
M$\,$92.  As in the VandenBerg et al study, the [Fe/H] values derived by
\citet{cbg09a} have been assumed.

Except at $M_{F606W} \gta 8$, where the solid curves deviate to the blue side
of the observed lower-MS stars, the isochrones do quite a good job of matching
the main sequences of the GCs over the entire range in [Fe/H] sampled by
them.  The biggest differences between theory and observations occur along the
lower RGB, where the models are too red.  However, the tendency of photometric
scatter due to blending to be preferentially blueward on the giant branch may
explain some fraction of such offsets (see \citealt{bs09}).
Curiously, the discrepancies resemble
the effect on the location of the RGB of varying the helium content: as shown
in Fig.~\ref{fig:fig7}, increasing $Y$ causes a larger temperature shift at the
base of the giant branch than near the tip.  On the other hand, it is possible
that our treatment of the atmospheric boundary condition is responsible for the
apparent difficulties (recall Fig.~\ref{fig:fig4}), or perhaps they signal some
problems with the color--$\teff$ relations that we have used or our treatment
of convection.  As noted in the introductory remarks given at the beginning
of \S~\ref{sec:obs}, predicted temperatures and colors are subject to many
uncertainties, and it should not be a surprise to find some discrepancies
between isochrones and observed CMDs.

To corroborate this point, we have generated isochrones for the case represented
in Fig.~\ref{fig:fig4} by the dot-dashed curve.  That is, a full set of
evolutionary tracks has been computed for [Fe/H] $= -2.40$, $Y = 0.25$, and
[$\alpha$/Fe] $= 0.4$ in which the surface boundary conditions have been derived
from the properties of MARCS model atmospheres at $\tau = 100$, with the small
increase in the pressure at that point implied by the corresponding Standard 
Solar Model.  As shown in Figure~\ref{fig:fig15}, the 13 Gyr isochrone derived
from these tracks, unlike the one plotted in the bottom, right-hand panel of
the previous figure, provides a good fit to the lower RGB stars of M$\,$92, but
not those at higher luminosities.  (Granted, the predicted turnoff is slightly
too blue, but an improved fit to the TO could be obtained, without affecting
the location of the lower RGB, simply by assuming a somewhat higher oxygen
abundance.)  In this example, the discrepancies along the upper giant branch
could be telling us that, e.g., our treatment of convection or the adopted
color--$\teff$\ relations are inadequate.   We could force the models to
provide an essentially perfect match to the data (by, for instance, suitable
adjustments of the color transformations or the atmospheric boundary
conditions), but the assumed distance and chemical abundances of M$\,$92 may
not be correct.  Although isochrones may need to be ``calibrated" for some
investigations, not doing so enables one to retain the predictive power of
stellar models.  In fact, it is remarkable that current stellar models perform
as well as they (appear to) do.

\section{Summary}
\label{sec:sum}

To obtain the correct understanding of stars and stellar populations, it is
important to determine the observed chemical abundances and, in the case of
complex systems (e.g., $\omega\,$Cen), their variations from star-to-star with
as much accuracy and detail as possible through spectroscopic and photometric
studies.  It is just as important to interpret such data using stellar models
for the observed chemistries because the most abundant metals (and helium) 
affect the predicted luminosities and temperatures of stars in different ways
(see, e.g., V12).  For this investigation, 126 grids of evolutionary tracks
have been computed for, in each case, masses from $0.12 \msol$ to a sufficiently
high mass that isochrones may be generated, using the accompanying software,
for ages $\gta 5$ Gyr, and arbitrary values of [Fe/H], $Y$, and [$\alpha$/Fe]
within the ranges $-2.4 \lta$ [Fe/H] $\lta +0.6$, $0.25 \le Y \le 0.33$, and
$-0.4 \lta$ [$\alpha$/Fe] $\lta +0.4$.  Comparisons of these computations with
the CMDs of M$\,$67, NGC$\,$6791, local field subdwarfs, and four GCs (47 Tuc,
M$\,$3, M$\,$5, and M$\,$92) provide encouraging support for the models.

One point worth additional emphasis is that our models (and the MARCS
color--$\teff$ relations) favor a relatively
hot temperature scale for metal-poor stars.  This
is not a new result, as virtually the same thing was found by \citet{bv01}.
Indeed, if anything, an even warmer $\teff$ scale would be implied by the use
of current MARCS model atmospheres as boundary conditions (see
Fig.~\ref{fig:fig4}).  Hotter stellar models could be at least part of the
explanation of the long-standing problem that isochrones applicable to GCs are
generally found to be slightly too red when well-supported estimates of the
cluster distances, reddenings, and chemical abundances are adopted (e.g., see
\citealt{van00}, \citealt{vbl13}, and our Fig.~\ref{fig:fig14}).  The difficulty
with this solution is that the same models appear to satisfy the subdwarf
constraint without needing any zero-point adjustment to the predicted colors
(see Fig.~\ref{fig:fig13}), though this could be a fortuitous result if
errors in the some of the subdwarf properties are compensating for the effects
of errors in other properties.  It is also possible that the apparent
inconsistencies occur because GC metallicities, as generally measured in 
bright giants, are not on the same scale as those for Population II dwarfs.
In particular, perhaps the [Fe/H] values of GCs are $\sim 0.15$--0.3 dex lower
than the majority of current estimates --- a possibility that is supported by
recent spectroscopic studies of M\,92 (\citealt{rs11}) and M\,15
(\citealt{pst06}, \citealt{sks11}), as well as other findings (\citealt{vbn14}).

Because of the overwhelming importance of oxygen for TO luminosity versus age
relations, the next paper in this series will provide extensive grids of
evolutionary tracks in which [O/Fe] is included among the chemical abundance
parameters that can be varied.  Among other things, Paper II will compare
predicted and observed luminosities of the RGB bump, which is known to be a
strong function of the oxygen abundance (see, e.g., \citealt{rc85}).  (Accurate
determinations of $V_{\rm bump}$ for more than 70 GCs are provided by
by \citealt{ngp13}.)  Following that investigation, fully consistent ZAHB
models will be presented for the grids reported here and in Paper II so that it
will be possible to assess their implications for distance determinations and
to interpret the observed colors of HB stars.

\acknowledgements
This paper has benefitted from a very careful and thoughtful critique by an
anonymous referee, whose important contributions to this paper are gratefully
acknowledged.  We thank Greg Feiden for providing the models that have been
compared with ours in Figure~\ref{fig:fig1}, Luca Casagrande for providing the
photometry that appeared in Figure~\ref{fig:fig10}, and Karsten Brogaard for
helpful comments on the metallicity of NGC$\,$6791.  DAV acknowledges the
support of a Discovery Grant from the Natural Sciences and Engineering Research
Council of Canada.  This research used the facilities of the Canadian Astronomy
Data Centre operated by the National Research Council of Canada with the support
of the Canadian Space Agency.
 
\appendix
\bigskip
\centerline{\bf APPENDIX}

All of the model grids that have been computed for this investigation may be
obtained from the {\it Canadian Advanced Network for Astronomical Research
(CANFAR)} web site,\footnote{http://www.canfar.phys.uvic.ca/vosui/\#/VRmodels}
together with several computer programs (in FORTRAN) that permit the user to
generate isochrones on the theoretical plane, to transpose the isochrones to
many different CMDs using the recent \citet{cv14} transformations, and to
calculate luminosity functions (LFs), isochrone population functions (IPFs),
and more.  The methods that we have developed over the years to facilitate
comparisons of models produced by the Victoria stellar evolutionary code with
observational data are well described in V12 and references therein.  In that
paper, we added the ability to interpolate within the canonical grids to create
grids of tracks with arbitrary helium abundances, $Y$, and/or metallicities,
[Fe/H], within the ranges $-3.0 \le$ [Fe/H] $\le -0.6$ and $0.25 \le Y \le
0.33$.  In this paper, we add the ability to interpolate the models in a third
chemical abundance parameter, either \afe\ (the elements O, Ne, Mg, Si, S, Ar,
Ca, and Ti as a group) or \mfe, where ${\rm m_i}$ refers to one of C, N, O, Ne,
Na, Mg, Al, Si, Ca, or Ti.  Although V12 used three-point interpolation for
both abundance parameters that they considered, we opted to employ linear
interpolation in $Y$ and [Fe/H] to make the scheme more robust (in the sense
that the age-mass relations which are critical for the isochrone interpolations
are guaranteed to remain monotonic) and more flexible to use.
Since only two values of \afe\ are represented at $\feh = -2.0$ and +0.6 in
the current computations (see Fig.~\ref{fig:fig5}), we also decided to use
linear interpolation for the third abundance parameter.

\section{Format of the Track Files}

As presented to the user, the evolutionary sequences are contained in EEP
(equivalent evolutionary phase) files which have been processed in such a
way that track points with the same model number are equivalent in every
track in every grid. Two caveats apply to this prescription. First, in grids
that extend to masses lower than $0.4 \msol$, tracks with masses $\leq
0.4 \msol$ are listed for equally logarithmically spaced ages from the zero-age
main sequence (ZAMS) point up to a maximum age of $\approx$ 30 Gyr. Second,
for those grids that contain tracks in which core contraction manifests itself
after core hydrogen exhaustion, the main sequence turnoff point EEP (MSTO)
becomes degenerate with the blue hook EEP (BLHK) for those lower mass tracks
in which the blue hook is not present.  We do not explictly list these
degenerate points: their presence (discussed below) is indicated by listing
the primary EEPS in the header lines for each track.

The canonical EEP file names (with the extension \verb+*.eep+) provide
all the abundance information for the tracks contained within them:
each one begins with a five character prefix that terminates with the
underscore character followed by three abundance specifications, e.g.,
\verb+a0zz_p4y29m18.eep+. In this example, \verb+a0+ indicates the
solar metals mixture (Asplund et al., 2009) and \verb+zz+ specifies
the entire group of $\alpha$-elements: decoding the rest of the name
from left to right, \verb+p4+ implies $\afe = +0.4$, \verb+y29+
implies $Y=0.29$, and \verb+m18+ implies $\feh=-1.8$. A grid of tracks
interpolated to $\afe=+0.2$, $Y=0.273$, and $\feh=-0.75$ would have
\verb+a0zz_p2y273m075.xeep+ as its file name, where the extension
\verb+.xeep+ distinguishes it from the canonical grids. Had the
interpolation been to $\afe=-0.2$, $Y=0.273$, and $\feh=+0.25$, the
file name would have been \verb+a0zz_m2y273p025.xeep+ --- that is, the
signs of the $\alpha$-element and iron abundances are denoted by either
\verb+p+ ($+$ve) or \verb+m+ ($-$ve).

For future reference with grids in which individual elements may be
enhanced differently with respect to some basic \afe\  abundance
ratio, such grids will have file names like \verb+a4xO_p1y25p02+. The
prefix {\tt a4xO} decodes as ``a basic $\afe=+0.4$ mixture with an
extra degree of enhancement of the element oxygen''. Decoding the
rest of the name, {\tt p1} means that oxygen has been incrementally
enhanced by +0.1 dex (above the amount in the basic $\afe$ mixture) so that
$\ofe=+0.5$, and {\tt y25p02} means $Y=0.25$ and $\feh=+0.2$. When the symbol
for an element consists of a single letter (like {\tt C}, {\tt N}, or {\tt O}),
that letter appears just before the underscore, and \verb+x+ is used as
a place-holder; otherwise, the third and fourth characters of the file
name give the two-letter symbol of the metal (e.g., {\tt Ne, Mg}) in
question.

The contents of a \verb+*.eep+ file are illustrated in Figure~\ref{fig:fig16}.
The header lines at the beginning of the file are reasonably straightforward
to interpret (note, in particular, that the assumed [$m$/Fe] values are
explicitly given for the main metals of interest),
but the header lines for individual tracks require some
explanation.  The columns labeled \verb+Match+, \verb+D(age)+, and
\verb+D(log Teff)+ are redundant: they list information about how the models
for the MS and SGB phases, which were obtained by solving the Lagrangian form
of the stellar structure equations, were matched (at the base of the RGB) to
the models for the subsequent evolution to the tip of the giant branch.  As
described in detail by \citet{van92}, a non-Lagrangian technique like the one
developed by \citet{egg71} was used to follow RGB evolution very efficiently.
The indicated age and $\teff$\ offsets were applied to the original track 
files for the RGB phase in order to obtain continuity with the Lagrangian
models.

The column labeled \verb+Zage+ lists the ZAMS age as we defined it in V12.
Six evolutionary phase points are listed under \verb+Primary EEPs+. The
default setting for each EEP-point is \verb+0+, which should be interpreted
to mean (except for the ZAMS) that that particular evolutionary phase does
not occur, or has not been reached, in that particular track. The ZAMS
EEP-point is listed as model \verb+1+ when the track is included in the full
isochrone interpolation scheme.  When listed as \verb+0+, it signals that an
interpolation to the isochrone age is to be made directly within that
track.  (Generally this applies to tracks with masses $\leq 0.4 \msol$.  Points
on the isochrone between those corresponding to these track masses are obtained
by spline interpolation.)  If the track evolves sufficiently, the MSTO
is listed at model \verb+801+, and if a blue hook occurs it is listed at model
\verb+921+. In the absence of a BLHK EEP, as is the case for the $0.9 \msol$
track in the grid shown, the base of the red giant branch (BRGB) occurs at
model \verb+1421+, the evolutionary pause on the giant branch (GBPS) at
model \verb+1621+, and the tip of the giant branch (GBTP) at model \verb+1921+.
When the BLHK is non-zero, as is the case for the tracks with masses
$>0.9 \msol$, the BRGB occurs at model \verb+1541+, the GBPS at \verb+1741+,
and the RGBT at \verb+2041+. 

In this example, tracks with masses $>0.9 \msol$ have BLHK EEPs while those
with masses $\leq 0.9 \msol$ do not.  (A ``nascent'' BLHK EEP may be identified
in some tracks to relax the interpolation scheme through the transition from
lower mass tracks with radiative cores at central H exhaustion to those higher
mass tracks with fully developed convective cores at the end of the MS phase;
see \citealt{bv01}.)  Consequently, 120 denerate EEP points would be inserted
between the MSTO and BRGB for the lower mass tracks that evolve at least as far
as the BRGB so that, for example, the BRGB in the $0.9 \msol$ track would
change from model \verb+1421+ to model \verb+1541+. The same approach works
when interpolating grids of tracks to some set of target abundances.

The first four columns in the listing for each track give the model
number, $\log L/L_{\sun}$, $\log T_{\rm eff}$, and the age in
Gyr. Columns 5 and 6 are reserved for the surface helium and \feh\
abundances (not implemented for the example shown in Fig.~\ref{fig:fig16},
while columns 7 and 8 list the derivatives of luminosity and of effective
temperature with respect to time.  The latter are needed for the calculation
of LFs and IPFs.

\section{Interpolation Software} 

The interpolation of isochrones, LFs, and/or IPFs
is made easy by the (FORTRAN) programs PBISO and PBIPF, which are improved
versions of the MKISO and MKIPF codes, respectively, that were presented 
in previous publications (see V12 and references therein).  We have now
developed a new program, PBMIX, that can be used to
interpolate within the canonical grids to produce a new grid of tracks at
some set of the (up to) three abundance parameters --- either ($\afe$, $Y$,
$\feh$) or ($\mfe$, $Y$, $\feh$) that is signaled by the file name prefix.
PBMIX makes use of a parameter file, PBMIX.PAR, that resides in the
working directory; it contains a listing of the track masses employed
in constructing the grid, the abundance ranges spanned by the grids,
and a listing of all the EEP files encompassed by the abundance ranges
specified. If PBMIX is run in a directory that doesn't contain a
PBMIX.PAR file, it will prompt you for all the requisite information and
construct one automatically. 

A sample file is shown in Figure~\ref{fig:fig17}. The first line lists the
prefix for the file names associated with the grids contained in the working
directory. (In this example, the grids have a basic $\afe = +0.4$
mixture with additional enhancements to $\ofe$.)  The next line begins
with the number of tracks that may be associated with each grid, followed by
the mass values --- since the masses are read in via a list-directed read
statement, they need only be separated by blanks and can be spread over several
lines (two lines in this example). The (fourth) line following the list of
masses gives the number of and canonical values for the abundance ratio $\ofe$.
(If the file prefix were \verb+a0zz+, this line would list the $\afe$ values
instead.) The next (fifth) line does the same thing for the helium abundances,
as does the (sixth) line for $\feh$.  Subsequently, the twenty-four canonical
file names derived from the file prefix and the tabulated abundances
--- each with a unique set of $\ofe$, $Y$, and $\feh$ --- are listed.

Running PBMIX is very simple. When PBMIX.PAR already exists, the
program simply prompts the user for the target $\afe$ value or the
$\Delta\mfe$ increment, and the target $Y$ and $\feh$ values, then it
creates the default output file name based on those targets and writes
the interpolated tracks to that file.  As in the case of the canonical
grids, isochrones may be generated for the interpolated grids for any
age in the range from $\sim 5$ Gyr to $\sim 15$ Gyr by executing PBISO.
The auxilliary code PBIPF may then be used to provide magnitudes and
colors in the Johnson-Cousins, 2MASS, Sloan, or {\it Hubble Space
Telescope} ACS or WFC3 photometric systems. (The color transformation
tables that are needed to accomplish this must be generated using the computer
programs and data provided by \citealt{cv14}.)  The same code provides the
option of producing LFs or IPFs.  A brief, but quite thorough, description
of how to use each code is provided in a manual that can also be retrieved
from the {\it CANFAR} web site.

\section{Tests of the Interpolation Software}

To demonstrate that the interpolation errors are small, even though linear
interpolation is employed for all three chemical abundance parameters, we have
computed a set of evolutionary tracks for values of [Fe/H] $= -0.1$,
[$\alpha$/Fe] $= +0.2$, and $Y = 0.27$ that are midway between their respective
grid values.  These particular abundances were chosen because they lie along the
[$\alpha$/Fe] versus [Fe/H] relationship that has been derived for stars in
the Galactic Bulge (e.g., \citealt[their Fig.~3]{rge10}), where the ``knee" in
that relation, which is believed to represent the point at which Type Ia
supernovae began to contribute to the chemical evolution of the Bulge, occurs
at an unusually high [Fe/H] value ($\approx -0.3$).  (To produce this grid, 
the necessary opacity data were generated in the usual way; see
\S~\ref{sec:abund}.)  It can be expected that this case will provide quite a
severe test of the interpolation scheme because (i) the difference between a
line that connects any two of the points that are fitted by a parabolic curve,
and that parabola, will be maximal at approximately the mid-point, and (ii) the
effects of opacity on mass-luminosity and age-mass relations are strongest at
the highest metal abundances.

However, as shown in Figure~\ref{fig:fig18}, isochrones derived from this set
of models differ only slightly from those for the same age that are obtained by
interpolation in the grids which are being provided for general use via this
paper.  At a value of $\log\,\teff$ that is 0.01 dex cooler than the turnoff
temperature, the subgiant branches of the interpolated isochrones are $\lta
0.04$ mag brighter than the SGBs of isochrones that have been derived from 
evolutionary tracks computed for the same values of $Y$, [$\alpha$/Fe], and
[Fe/H].  This is not negligible, but as noted above, the interpolation errors
are expected to be larger for this case than for most other choices of the
abundance parameters that are representative of the observed abundances in
stars.  For instance, the transition from [$\alpha$/Fe] $\approx +0.4$ to 0.0
typically begins at [Fe/H] $< -1.0$ in dwarf galaxies (\citealt{vis04}), and
the second test case that we have considered (for $Y=0.26$, [$\alpha$/Fe]
$= +0.2$, and [Fe/H] $= -0.9$) shows that the computed and interpolated
isochrones which are applicable to such systems will be indistinguishable (see
Fig.~\ref{fig:fig18}).

\begin{figure}
\plotone{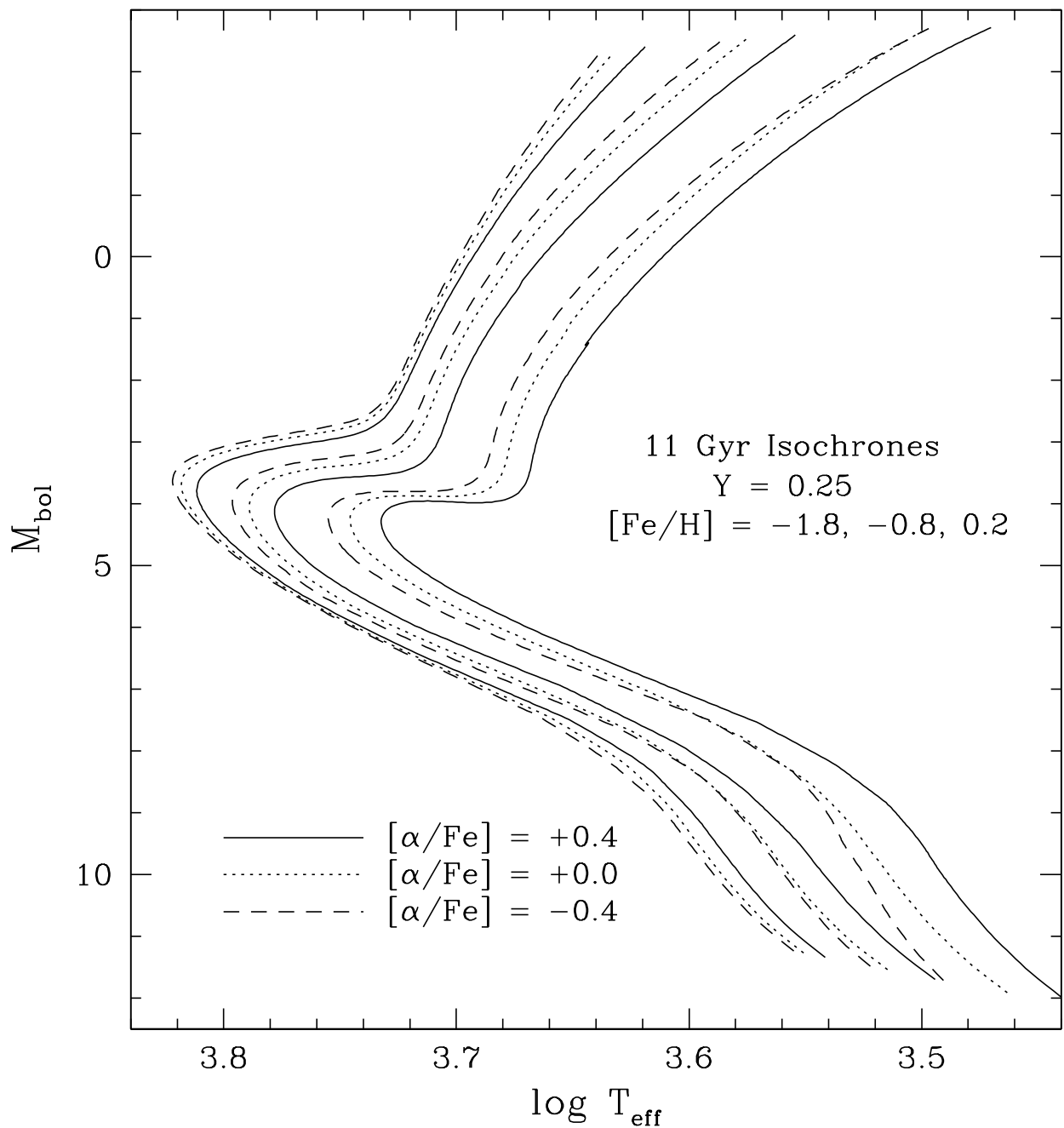}
\caption{Similar to the previous figure, except that the dependence of 11 Gyr
isochrones for $Y = 0.25$ and [Fe/H] $= -2.0$, $-1.0$, and 0.0 on [$\alpha$/Fe]
is shown.  As indicated, the solid, dotted, and dashed curves
assume [$\alpha$/Fe] $= +0.4$, 0.0, and $-0.4$, respectively.}
\label{fig:fig8}
\end{figure}

\begin{figure}
\plotone{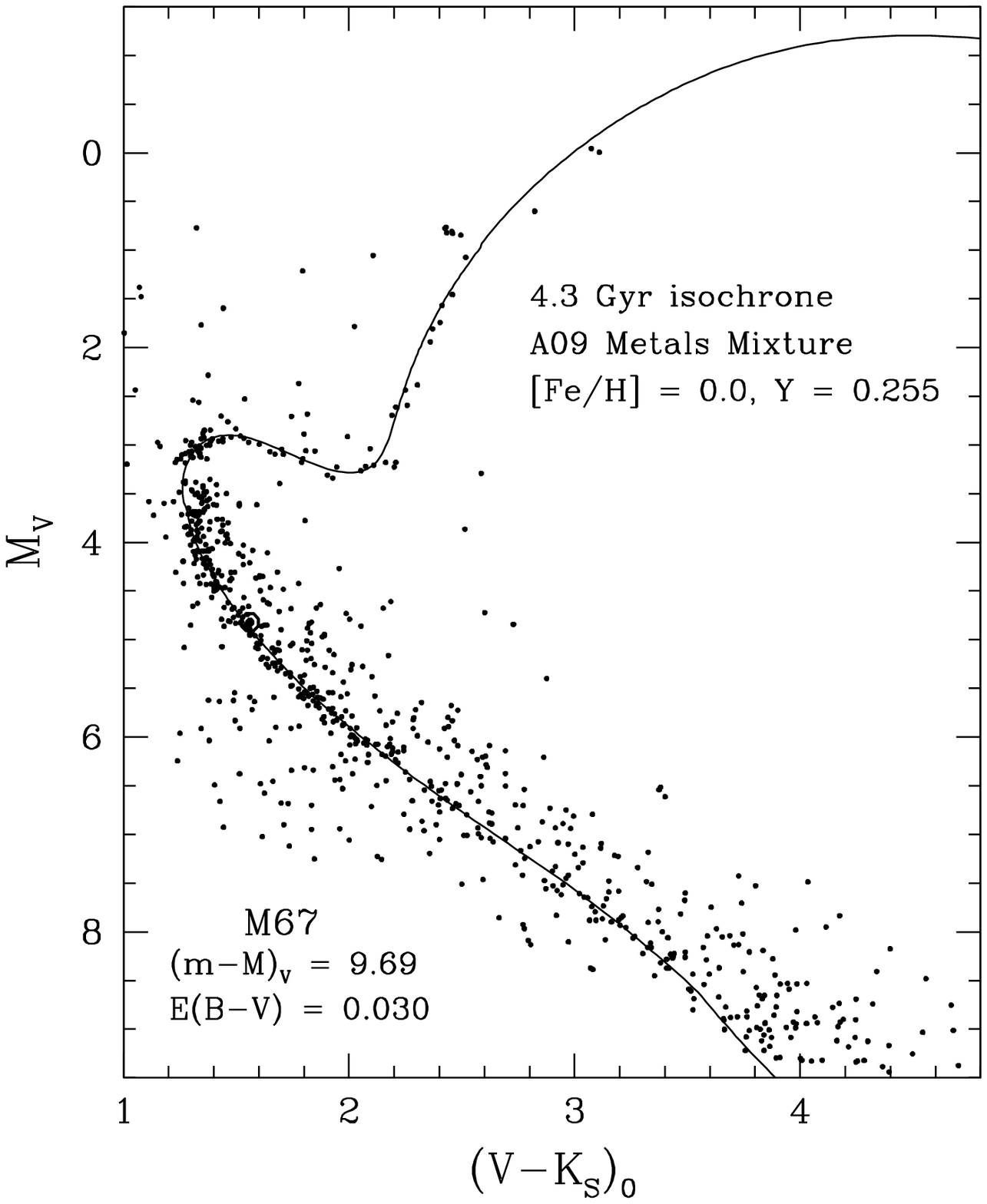}
\caption{Overlay of a 4.3 Gyr isochrone for the indicated values of [Fe/H] and
$Y$ onto the CMD of M$\,$67 (see \citealt{bsv10}) on the assumption of 
$E(B-V) = 0.03$ (from SF11), $E(V-K_S) = 2.76\,E(B-V)$ (\citealt{cv14}) and an
apparent distance modulus $(m-M)_V = 9.69$.  The solar symbol (at $V-K = 1.56$,
$M_V = 4.82$) indicates the location of the Sun on this diagram.  The main point
of this plot is that the observed MS and RGB are matched quite well by the
models in both an absolute and systematic sense.}
\label{fig:fig9}
\end{figure}

\begin{figure}
\plotone{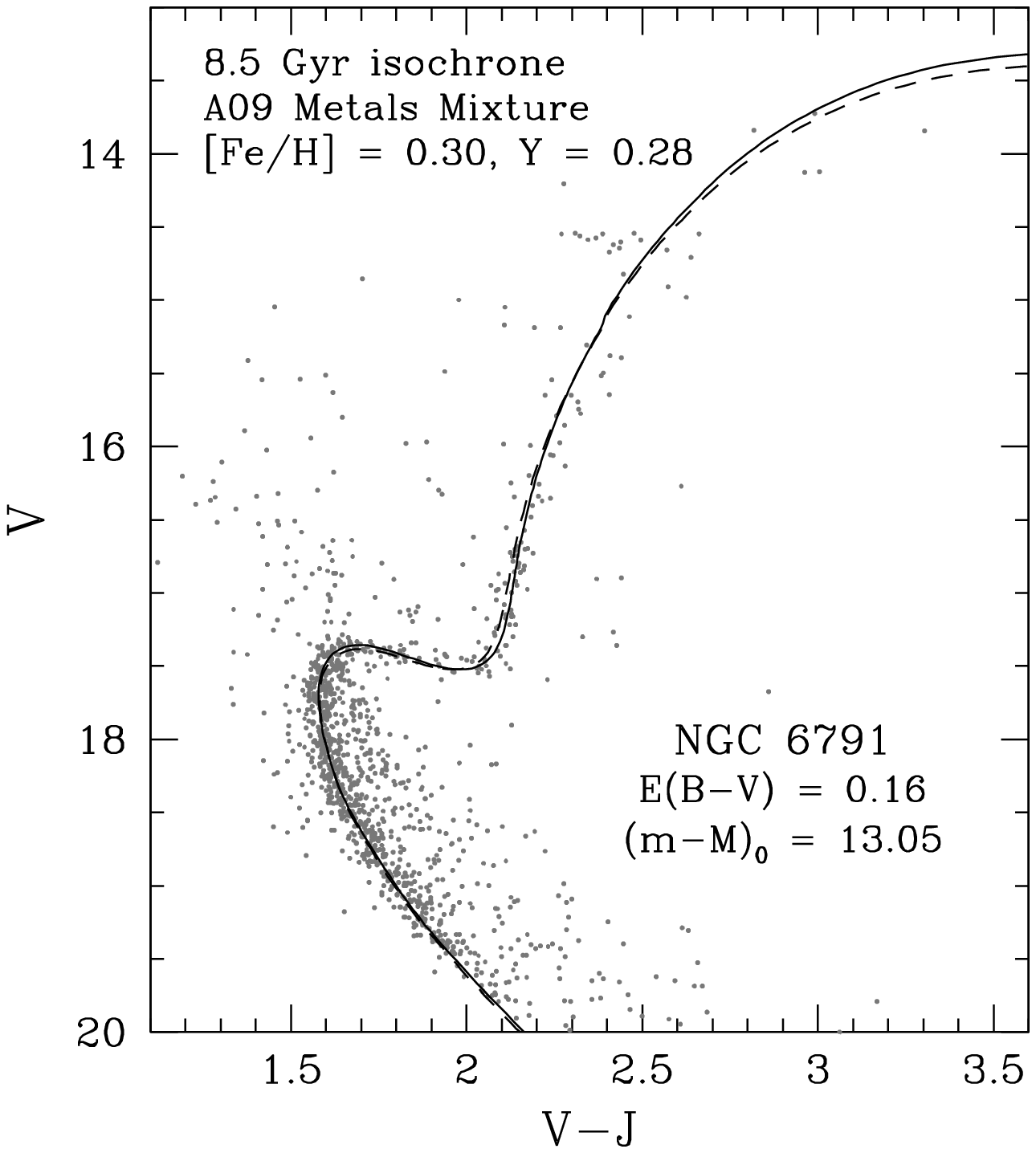}
\caption{Using the reddening-corrected transformations given by \citet{cv14},
an isochrone for the indicated age and chemical abundances (solid curve) has
been overlaid onto $V\,J$ photometry of NGC$\,$6791 (\citealt{bsv10}).  For
reasons discussed in the text, the observed colors have been adjusted to the
red by 0.04 mag.  The dashed curve represents an 8.0 Gyr isochrone for
$Y = 0.30$ and the \citet{gs98} metal abundances, scaled to [Fe/H] $= +0.35$.
Both isochrones were transposed to the observed plane assuming the reddening
and true distance modulus specified in the lower right-hand corner.}
\label{fig:fig10}
\end{figure}

\begin{figure}
\plotone{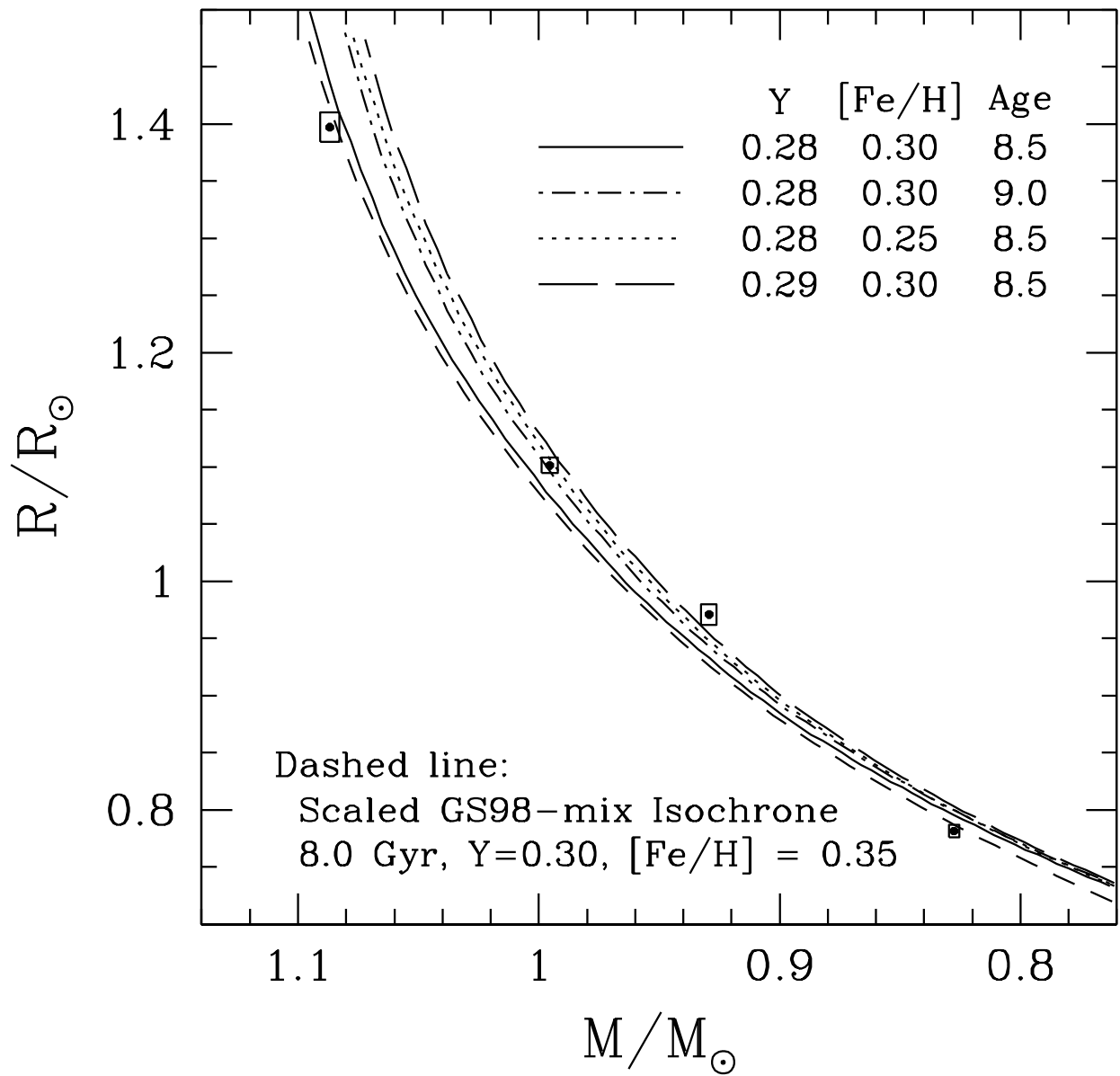}
\caption{Comparison on the mass-radius plane of isochrones for the indicated
values of $Y$, [Fe/H], and age with the properties of the binaries V18 and
V20 in NGC$\,$6791 (as represented by the small {\it filled circles} and the
$1\,\sigma$ error rectangles).  Numerical values for the latter are given by
\citet[see their Table 1]{bvb12}.  The solid and dashed curves represent
the same isochrones that were plotted in the previous figure.}
\label{fig:fig11}
\end{figure}

\begin{figure}
\plotone{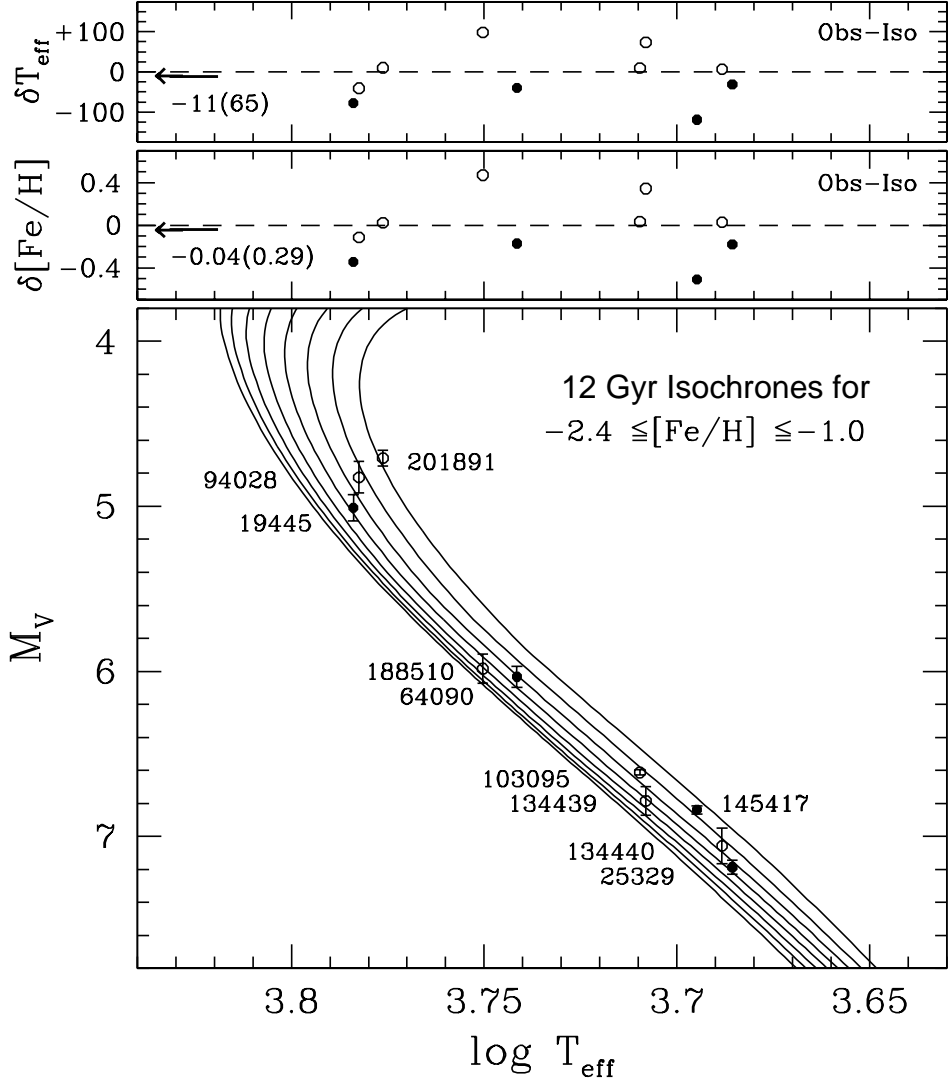}
\caption{{\it Lower panel}: superposition of those nearby subdwarfs with [Fe/H]
$\lta -1.0$ that have the most precise $M_V$ values (from {\it Hipparcos}) onto
12 Gyr isochrones for the indicated [Fe/H] values (in 0.2 dex increments).
Open circles denote subdwarfs with [Fe/H] $\ge -1.5$, whereas the stars with
lower metallicities have been plotted as filled circles.  The stars are 
identified by their HD numbers.  See the text for the sources of their $\teff$
and [Fe/H] values.  {\it Middle panel}: the difference between the observed
[Fe/H] value and that inferred for each star from the interpolated (or
extrapolated) isochrone that matches its location in the bottom panel.
{\it Upper panel}: the shift in $\teff$\ that would have to be applied to each
subdwarf in order for it to be located on the isochrone (in the bottom panel)
for its observed [Fe/H] value.  The arrows and adjacent numbers in the middle
and upper panels indicate the mean values of $\delta\,$[Fe/H] and $\delta\teff$,
respectively, along with the standard deviations (in parentheses) for the
subdwarf samples in the sense ``observed minus predicted".}
\label{fig:fig12}
\end{figure}

\begin{figure}
\plotone{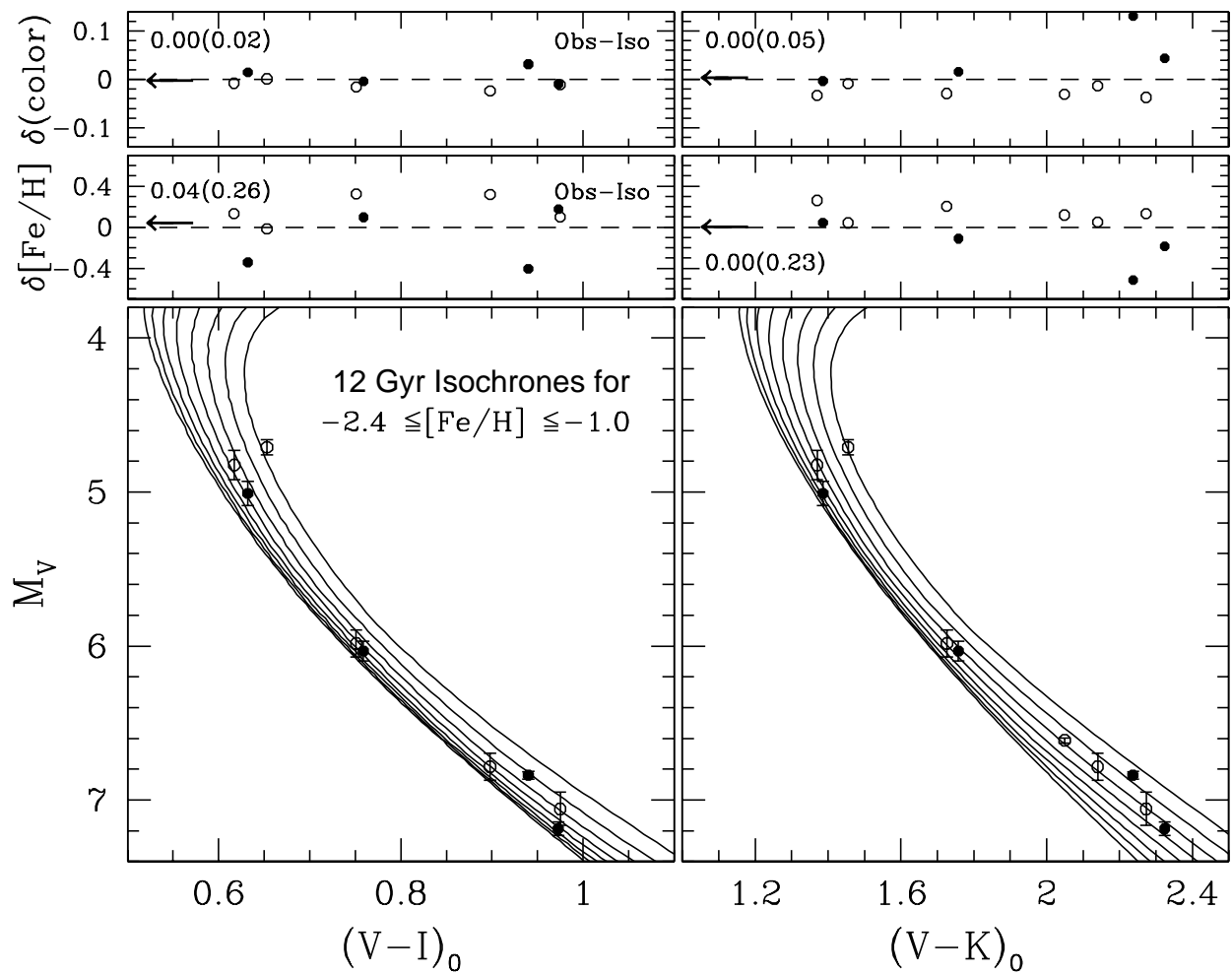}
\caption{Similar to the previous figure, except that the properties of the same
subdwarfs are compared with isochrones that have been transposed to the $V-I$
and $V-K$ color planes using the MARCS color--$\teff$\ relations reported by
\citet{cv14}.  The observed colors are from the study by \citet{crm10}.}
\label{fig:fig13}
\end{figure}

\begin{figure}
\plotone{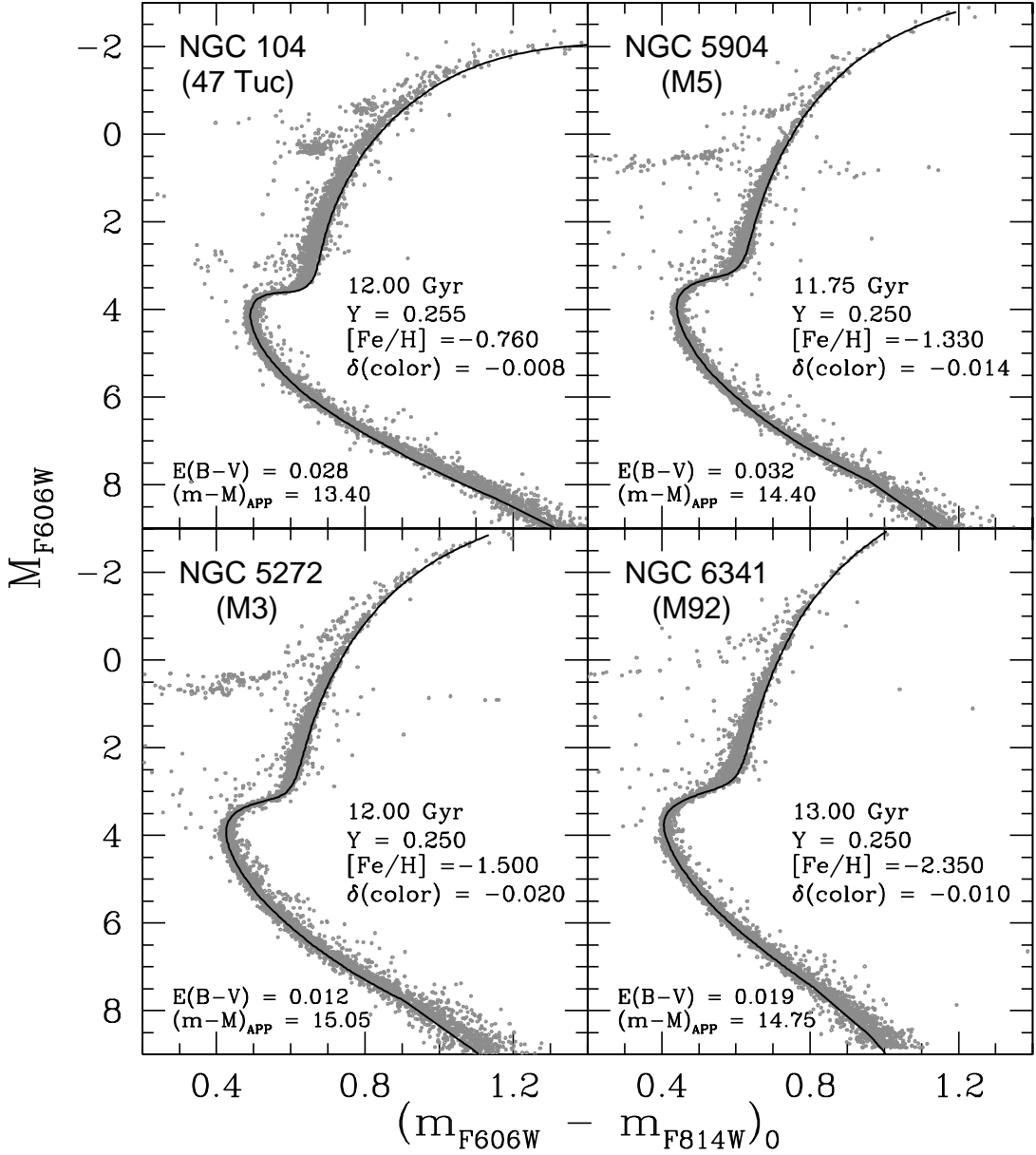}
\caption{Fits of isochrones for the indicated ages and chemical abundances to
the CMDs of 47 Tuc, M$\,$3, M$\,$5, and M$\,$92 after the latter have been
adjusted in the vertical and horizontal directions by the indicated distance
moduli and reddenings (from the SF11 dust maps, and assuming $E(m_{F606W} -
m_{F814W}) = 0.997\,E(B-V)$; \citealt{cv14}), respectively.  In order for the
selected isochrones to match the intrinsic turnoff colors, the model loci had
to be corrected by the $\delta$(color) amounts specified in each panel.  The
source of the cluster photometry is \citet{sbc07}.}
\label{fig:fig14}
\end{figure}

\begin{figure}
\plotone{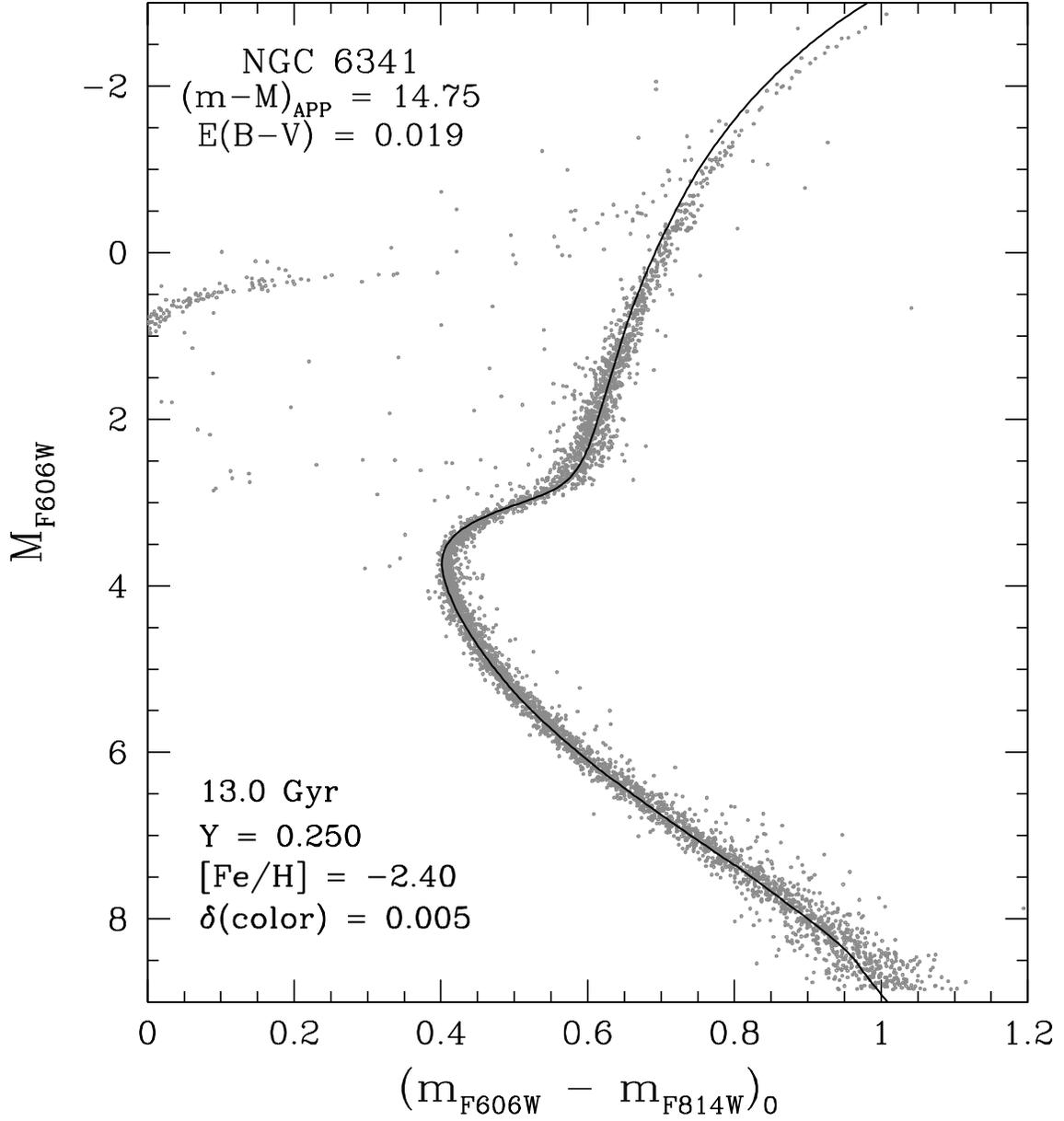}
\caption{As in the bottom right-hand panel of the previous figure, except that
the tracks on which the isochrones are based were obtained by matching MARCS
model atmospheres to the interior structures at $\tau = 100$ and the pressure
at that depth was increased by $\delta\log P = +0.042$ (see
Fig.~\ref{fig:fig4}).  Note that a slightly lower [Fe/H] values was assumed in
the one set of models that was computed for this case and that a small redward
correction was applied to the isochrone colors in order to provide the best
fit to the cluster MS stars.}
\label{fig:fig15}
\end{figure}

\begin{figure}
\figurenum{16}
\epsscale{0.8}
\plotone{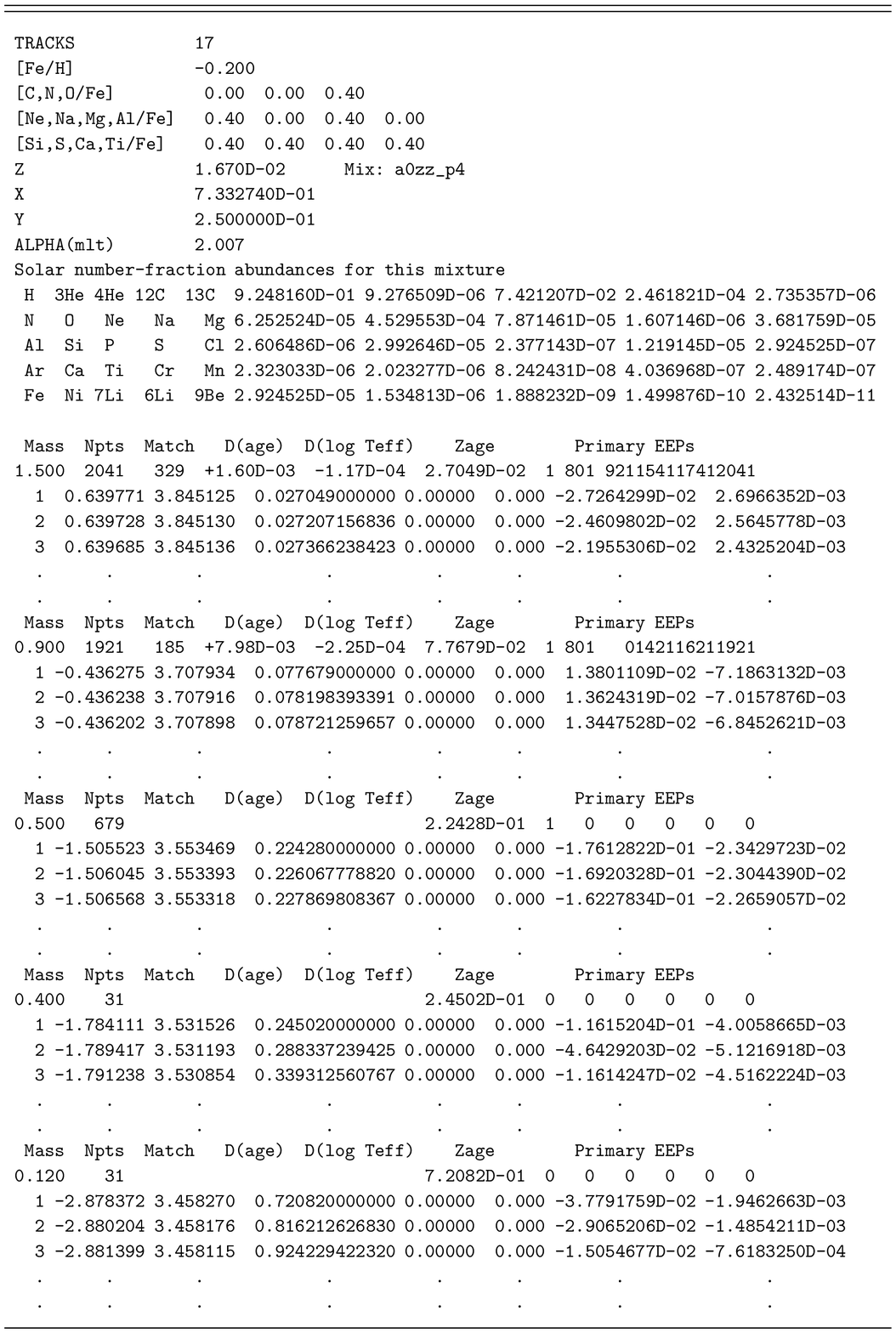}
\caption{Header lines and entries for selected tracks in the file
{\tt a0zz{\usebox{\uscorebox}}p4y25m02.eep}. Listings of the primary
EEPs for each track provide the information needed to connect the
secondary EEPs between tracks of {\it different masses within a grid}
for isochrone interpolation, and between tracks with {\it identical
masses in different grids} for track interpolation.}
\label{fig:fig16}
\end{figure}

\begin{figure}
\figurenum{17}
\epsscale{0.5}
\plotone{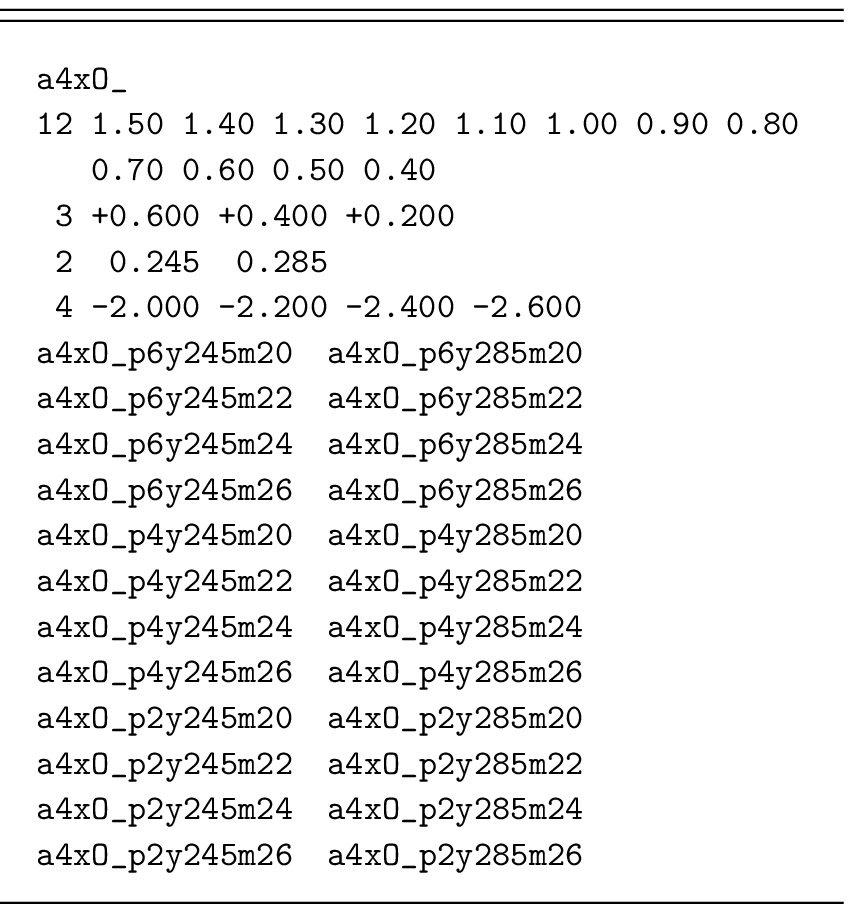}
\caption{A sample PBMIX.PAR file for a set of twenty-four canonical
grids containing up to twelve tracks with masses ranging from 0.4 to
$1.5 \msol$, constructed with a basic $\afe=+0.4$ mixture at
combinations of $\ofe[+0.2, +0.4, +0.6]$, $Y[0.245, 0.285]$, and
$\feh[-2.0, -2.2, -2.4, -2.6]$.}
\label{fig:fig17}
\end{figure}

\begin{figure}
\figurenum{18}
\epsscale{0.8}
\plotone{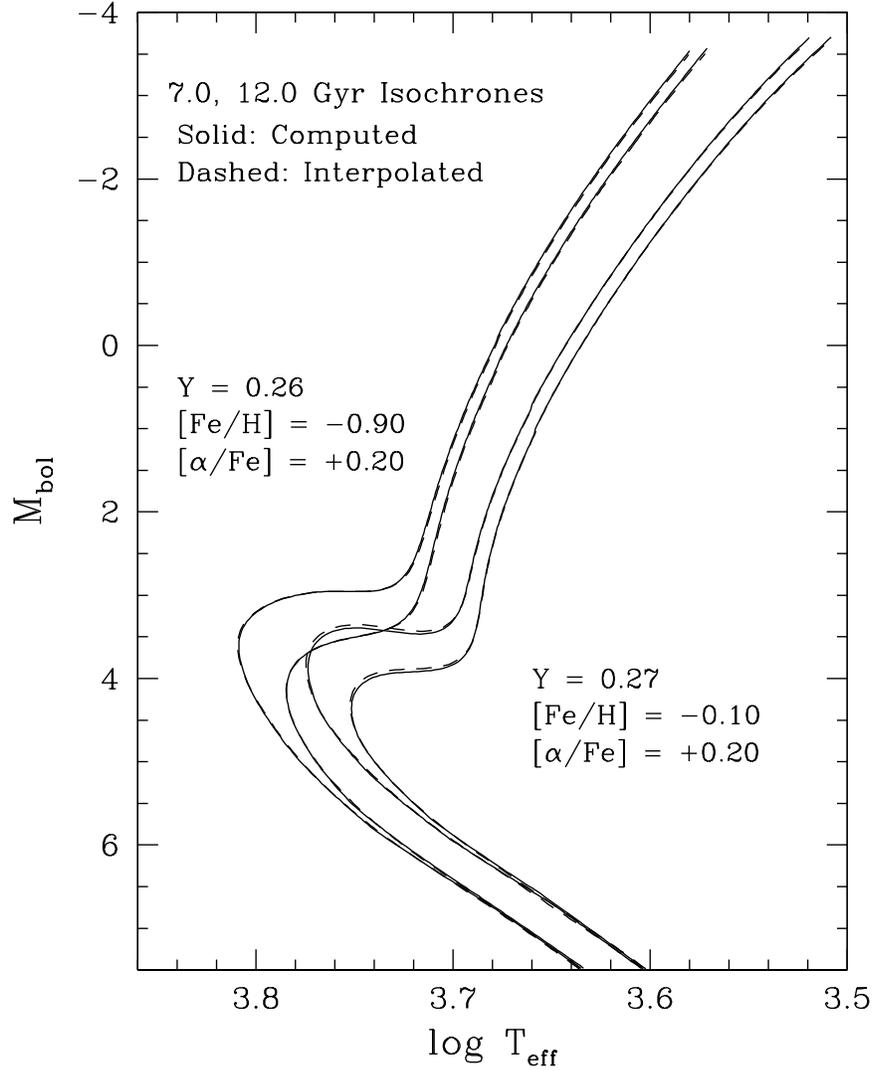}
\caption{Comparisons of 7.0 and 12.0 Gyr isochrones for the indicated values
of $Y$, [Fe/H], and [$\alpha$/Fe].  The dashed curves were obtained by
interpolating in the grids of models that are presented in this paper.  The
solid curves are based on grids of evolutionary tracks that were computed
for the specified abundances, and thus do not involve any interpolations
whatsoever of the chemical abundance parameters.}
\label{fig:fig18}
\end{figure}

We would have preferred to compute grids of evolutionary sequences for
$\delta$[$\alpha$/Fe] $= 0.2$, but the model atmospheres that are needed to
provide the surface boundary conditions for the lowest mass models are not
currently available for such a fine spacing of this quantity.  It turns out, in
fact, that most of the interpolation errors at high metallicities occur because
a grid spacing of 0.4 dex range in [$\alpha$/Fe] is too large.  We reached this
conclusion after comparing computed and interpolated isochrones for a third
case; specifically, [$\alpha$/Fe] $= 0.0$ (one of the grid values), $Y = 0.27$,
and [Fe/H] $= +0.30$, where the latter quantities are at the mid-points of the
grid values, $Y = 0.25$, 0.29 and [Fe/H] $= +0.2$, $+0.4$, respectively.  To
avoid making Fig.~\ref{fig:fig8} too complicated, we elected not to plot these
two isochrones, but we found that they superimpose each other so well that the
differences between them are barely discernible.  

V12 (see their Figs.~4--6) had previously demonstrated that the errors
associated with $Y$, [Fe/H] interpolations are negligible if the grid spacings
of these variables are $\delta\,Y = 0.04$ and $\delta$[Fe/H] $= 0.2$ dex.
Although they used three-point interpolations in the tests that they conducted,
we have found that linear interpolations work nearly as well (even at high
metal abundances).  Thus, there appears to be only a small regime of parameter
space ([$\alpha$/Fe] roughly halfway between the grid values, but only at
relatively high [Fe/H] values) where minor interpolation errors can be expected.
Elsewhere, such errors are of no consequence whatsoever.


\clearpage

\clearpage

\clearpage

\clearpage

\clearpage

\clearpage

\clearpage

\clearpage

\clearpage

\clearpage

\clearpage

\clearpage
 
\clearpage
 
\clearpage

\clearpage

\clearpage

\begin{lrbox}{\uscorebox}
\verb+_+
\end{lrbox}

\clearpage

\clearpage

\clearpage
\begin{deluxetable}{rccc}
\tabletypesize{\footnotesize}
\tablewidth{400pt}
\tablecaption{Adopted ($\log\,N$) Metal Abundances\tablenotemark{a}
 \label{tab:tab1}}
\tablewidth{0pt}
\tablehead{\colhead{Element} & \colhead{[$\alpha$/Fe] $= 0.0$
 Mix\tablenotemark{b}} & \colhead{[$\alpha$/Fe] $= -0.4$ Mix} &
 \colhead{[$\alpha$/Fe] $= +0.4$ Mix} }
\startdata
\noalign{\vskip 3pt}
  C\phm{ent} & 8.43 & 8.43 & 8.43 \\
  N\phm{ent} & 7.83 & 7.83 & 7.83 \\
  O\phm{ent} & 8.69 & 8.29 & 9.09 \\
 Ne\phm{ent} & 7.93 & 7.53 & 8.33 \\
 Na\phm{ent} & 6.24 & 6.24 & 6.24 \\
 Mg\phm{ent} & 7.60 & 7.20 & 8.00 \\
 Al\phm{ent} & 6.45 & 6.45 & 6.45 \\
 Si\phm{ent} & 7.51 & 7.11 & 7.91 \\
  P\phm{ent} & 5.41 & 5.41 & 5.41 \\
  S\phm{ent} & 7.12 & 6.72 & 7.52 \\
 Cl\phm{ent} & 5.50 & 5.50 & 5.50 \\
 Ar\phm{ent} & 6.40 & 6.00 & 6.80 \\
  K\phm{ent} & 5.03 & 5.03 & 5.03 \\
 Ca\phm{ent} & 6.34 & 5.94 & 6.74 \\ 
 Ti\phm{ent} & 4.95 & 4.55 & 5.35 \\
 Cr\phm{ent} & 5.64 & 5.64 & 5.64 \\
 Mn\phm{ent} & 5.43 & 5.43 & 5.43 \\
 Fe\phm{ent} & 7.50 & 7.50 & 7.50 \\
 Ni\phm{ent} & 6.22 & 6.22 & 6.22 \\
\enddata
\tablenotetext{a}{Assuming the scale in which $\log\,N($H$) =12.0$.}
\tablenotetext{b}{Solar abundances from Asplund et al.~(2009).}
\end{deluxetable}

\end{document}